\documentclass[12pt]{book}

\usepackage{amsmath,amsfonts,amsthm,amssymb}
\usepackage{graphicx}
\usepackage{color}
\usepackage{subfigure}
\usepackage{cite}
\usepackage{dcolumn}
\usepackage{bm}
\usepackage{book tabs}
\usepackage{braket}
\usepackage{setspace}
\newcounter{one}
\newcommand{\kB}{{k_\mathrm{B}}}
\newcommand{\TR}{{T_\mathrm{R}}}
\newcommand{\TL}{{T_\mathrm{L}}}
\newcommand{\QR}{{Q_\mathrm{R}}}
\newcommand{\QL}{{Q_\mathrm{L}}}
\newcommand{\UR}{{U_\mathrm{R}}}
\newcommand{\UL}{{U_\mathrm{L}}}

\newcommand{\WL}{{W_\mathrm{L}}}
\newcommand{\NR}{{N_\mathrm{R}}}
\newcommand{\NL}{{N_\mathrm{L}}}
\newcommand{\betaR}{{\beta_\mathrm{R}}}
\newcommand{\betaL}{{\beta_\mathrm{L}}}
\newcommand{\fR}{{f_\mathrm{R}}}
\newcommand{\fL}{{f_\mathrm{L}}}
\newcommand{\chicR}{{\chi_{c\mathrm{R}}}}
\newcommand{\chicL}{{\chi_{c\mathrm{L}}}}
\newcommand{\chihR}{{\chi_{h\mathrm{R}}}}
\newcommand{\chihL}{{\chi_{h\mathrm{L}}}}
\newcommand{\chiQL}{{\chi_{Q}^{\mathrm{L}}}}

\newcommand{\chiNL}{{\chi_{N}^{\mathrm{L}}}}

\newcommand{\chiqR}{{\chi_{q}^{\mathrm{R}}}}
\newcommand{\chiqL}{{\chi_{q}^{\mathrm{L}}}}

\newcommand{\AcR}{{A_{c\mathrm{R}}}}
\newcommand{\AcL}{{A_{c\mathrm{L}}}}
\newcommand{\AhR}{{A_{h\mathrm{R}}}}
\newcommand{\AhL}{{A_{h\mathrm{L}}}}
\newcommand{\ANR}{{A_N^{\mathrm{R}}}}
\newcommand{\ANL}{{A_N^{\mathrm{L}}}}

\newcommand{\AQR}{{A_Q^{\mathrm{R}}}}
\newcommand{\AQL}{{A_Q^{\mathrm{L}}}}

\newcommand{\AqR}{{A_q^{\mathrm{R}}}}
\newcommand{\AqL}{{A_q^{\mathrm{L}}}}

\newcommand{\Ah}{{A_h}}
\newcommand{\Ac}{{A_c}}

\newcommand{\muR}{{\mu_\mathrm{R}}}
\newcommand{\muL}{{\mu_\mathrm{L}}}
\newcommand{\EG}{{E_\mathrm{G}}}
\newcommand{\JQR}{{J_Q^{\mathrm{R}}}}
\newcommand{\JQL}{{J_Q^{\mathrm{L}}}}

\newcommand{\SRdot}{{\dot{S}_\mathrm{R}}}
\newcommand{\SLdot}{{\dot{S}_\mathrm{L}}}
\newcommand{\QRdot}{{\dot{Q}_\mathrm{R}}}
\newcommand{\QLdot}{{\dot{Q}_\mathrm{L}}}

\newcommand{\Sdot}{{\dot{S}}}
\newcommand{\Dl}{{D_{\ell}}}

\newcommand{\ain}{{a^{\text{in}}}}
\newcommand{\aout}{{a^{\text{out}}}}
\newcommand{\bin}{{b^{\text{in}}}}
\newcommand{\bout}{{b^{\text{out}}}}
\newcommand{\anin}{{a_n^{\text{in}}}}
\newcommand{\anout}{{a_n^{\text{out}}}}
\newcommand{\bnin}{{b_n^{\text{in}}}}
\newcommand{\bnout}{{b_n^{\text{out}}}}
\newcommand{\vecain}{{\bm{a}^{\text{in}}}}
\newcommand{\vecaout}{{\bm{a}^{\text{out}}}}
\newcommand{\vecbin}{{\bm{b}^{\text{in}}}}
\newcommand{\vecbout}{{\bm{b}^{\text{out}}}}
\newcommand{\SLL}{{S_\text{LL}}}
\newcommand{\SLR}{{S_\text{LR}}}
\newcommand{\SRL}{{S_\text{RL}}}
\newcommand{\SRR}{{S_\text{RR}}}
\newcommand{\SLa}{{S_{\text{L}\alpha}}}
\newcommand{\SLb}{{S_{\text{L}\beta}}}
\newcommand{\IL}{{I_{\mathrm{L}}}}
\newcommand{\cre}[2]{{{#1}_{{#2}}^{\dagger}}}
\newcommand{\ani}[2]{{{#1}_{{#2}}}}
\newcommand{\dcre}[3]{{{#1}_{{#2},{#3}}^{\dagger}}}
\newcommand{\dani}[3]{{{#1}_{{#2},{#3}}}}
\newcommand{\dlcre}[3]{{{#1}_{\text{{#2}},{#3}}^{\dagger}}}
\newcommand{\dlani}[3]{{{#1}_{\text{{#2}},{#3}}}}
\newcommand{\dagg}[1]{{{#1}^\dagger}}
\newcommand{\conjg}[1]{{{#1}^\ast}}
\newcommand{\Tr}[1]{{\text{Tr}[{#1}]}}
\newcommand{\expect}[1]{{\langle{#1}\rangle}}
\newcommand{\cumulant}[1]{{\langle\!\langle{#1}\rangle\!\rangle}}
\newcommand{\deriv}[2]{\frac{\text{d}{#1}}{\text{d}{#2}}}

\begin{document}

\title{Thermodynamics of Mesoscopic Quantum Systems}
\author{Kaoru Yamamoto \\ Department of Physics, The University of Tokyo }
\date{January 5, 2015}
\maketitle

\chapter*{Acknowledgements}\label{acknowledge}
\addcontentsline{toc}{chapter}{Acknowledgements}
It is a pleasure to thank Professor Naomichi Hatano for his valuable suggestions and comments. The author is also grateful to Eiki Iyoda, Professor Kensuke Kobayashi and the members of his group, Takashi Mori, Professor Keiji Saito, Tatsuhiko Shirai for fruitful discussions and comments.  
Finally, I appreciate the members of the Hatano group for their tender supports.

\clearpage

\begin{center}
\large{{\bf Abstract}}
\end{center}

In the present thesis, we study the heat flow in mesoscopic one-dimensional transport systems.
Using the analysis of full counting statistics, we calculate the cumulant generating function of the particle and heat flows and prove its symmetry. The symmetry produces the relations among transport coefficients of the particle and heat flows when we expand these flows with respect to the appropriate affinities. 
Moreover, we consider the generalized flows which are superpositions of the particle and energy flows. 
We show that we can choose the appropriate affinities of the generalized flows and derive the relations among their transport coefficients when we expand the generalized flows with respect to their affinities. 

\tableofcontents

\chapter{Intoroduction and Notation} \label{intro}
In this chapter, we explain a brief history of studies on linear irreversible thermodynamics and its application to thermoelectric devices. 
We then introduce mesoscopic transport systems, in particular mesoscopic thermoelectric systems, which can be beyond linear-response regime.  
We finally explain notations which are used throughout this thesis.
\section{Introduction}
\subsection{Thermoelectric device as a heat engine and linear irreversible thermodynamics}
Thermoelectric devices, which convert heat to work or vice versa, have helped the development of our society. Applications of thermoelectric devices include thermoelectric generator, thermoelectric refrigerator, and so on \cite{goupil2011}.
In order to analyze thermoelectric devices, linear irreversible thermodynamics has been used \cite{goupil2011,callen}. Linear irreversible thermodynamics is a phenomenological formalism which was constructed mainly by Onsager. Although the standard thermodynamics can treat only equilibrium, not transport problems, the linear irreversible thermodynamics lets us handle the latter.

\begin{figure}
\begin{center}
\includegraphics[width=6.5cm]{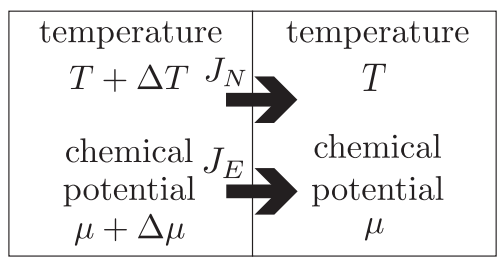}
\end{center}
\caption{The system in consideration.}
\label{fig:1system}
\end{figure}
Let us overview Onsager's formalism of linear irreversible thermodynamics \cite{goupil2011,callen,onsager1931a}. We consider two regions whose temperatures and chemical potentials we can define as shown in Fig.~\ref{fig:1system}. 
We then assume that the system has reached a non-equilibrium steady state  in which there are constant flows from one region to the other. 
Let us expand the particle and energy flows, $J_N$ and $J_E$, in terms of $\mu/T$ and $-1/T$:
\begin{align}
J_N &= L_{NN}\Delta\!\!\left(\frac{\mu}{T}\right) + L_{NE}\Delta\!\!\left(-\frac{1}{T}\right), \\
J_E &= L_{EN}\Delta\!\!\left(\frac{\mu}{T}\right) + L_{EE}\Delta\!\!\left(-\frac{1}{T}\right).
\end{align}
The expansion coefficients $L_{NN}$, $L_{NE}$, $L_{EN}$, and $L_{EE}$ are the transport coefficients for the particle and energy flows. The Onsager-Casimir relations are the relations among them under a magnetic field $B$ \cite{onsager1931a,onsager1931b,casimir1945}:
\begin{align}
L_{NN}(B) &= L_{NN}(-B), \label{eq:LNN} \\
L_{NE}(B) &= L_{EN}(-B), \label{eq:LNE}\\
L_{EE}(B) &= L_{EE}(-B). \label{eq:LEE}
\end{align}

Recently, the thermoelectric device has been studied as a heat engine from a point of view of the efficiency at the maximum power \cite{vandenboreck2005,saito2010,benenti2011,saito2011,brandner2013,brandner2013multi,Benenti}. 
Let us describe the setup to consider the thermoelectric device as a heat engine.
We set the chemical potential of the right reservoir to be higher than the left, while the temperature of the left reservoir to be higher than the right so that an electric current may go from left to right against the difference of the chemical potential; in other words, we set $\Delta T>0$ and $\Delta \mu <0$ in Fig.~\ref{fig:1system}. 
What happens per unit time is the following. Electrons gain heat defined by $J_Q=J_E-\mu J_N$ from the hot left reservoir, go to the right against the potential difference $-\Delta\mu = |\Delta \mu|$, during which electrons do the work of amount $J_N|\Delta\mu|$. 
We can thus consider this system as a heat engine. 
The efficiency  $\eta$ of this engine is therefore given by
\begin{equation}
\eta = \frac{J_N|\Delta\mu|}{J_Q}. \label{eq:1efficiency}
\end{equation}

In order to analyze the efficiency of a heat engine, linear irreversible thermodynamics is still useful. 
When the relations for the particle and energy flows, Eqs.~\eqref{eq:LNN}--\eqref{eq:LEE}, are valid, we can prove that the Onsager-Casimir relations are also valid for the coefficients of the particle and \textit{heat} flows \cite{callen}. 
In order to prove it, we expand $J_N$ and $J_Q$ as follows:
\begin{align}
J_N &= L_{NN}\Delta\!\!\left(\frac{\mu}{T}\right) + L_{NE}\Delta\!\!\left(-\frac{1}{T}\right) \notag \\
 &= L_{NN} \left[\frac{\Delta\mu}{T}+\mu\Delta\!\!\left(\frac{1}{T}\right) \right] + L_{NE}\Delta\!\!\left(-\frac{1}{T}\right) \notag \\
 &= L_{NN}\frac{\Delta\mu}{T} + (L_{NE}-\mu L_{NN})\Delta\!\!\left(-\frac{1}{T}\right) \label{eq:JNLL},
\end{align}
and
\begin{align}
J_Q &= J_E-\mu J_N \notag \\
 &= (L_{EN}-\mu L_{NN})\Delta\!\!\left(\frac{\mu}{T}\right) + (L_{EE}-\mu L_{NN})\Delta\!\!\left(-\frac{1}{T}\right) \notag \\
 &=  (L_{EN}-\mu L_{NN})\left[\frac{\Delta\mu}{T}+\mu\Delta\!\!\left(\frac{1}{T}\right) \right]+ (L_{EE}-\mu L_{NN})\Delta\!\!\left(-\frac{1}{T}\right) \notag \\
 &= (L_{EN}-\mu L_{NN})\frac{\Delta\mu}{T} +[L_{EE}-\mu(L_{EN}+L_{NE})+\mu^2L_{NN}]\Delta\!\!\left(-\frac{1}{T}\right). \label{eq:JQLL}
\end{align}
Let us define new transport coefficients as follows:
\begin{align}
J_N &= G_{NN}\frac{\Delta\mu}{T} + G_{NQ}\Delta\!\!\left(-\frac{1}{T}\right), \label{eq:JNGG}\\
J_Q &= G_{QN}\frac{\Delta\mu}{T} + G_{QQ}\Delta\!\!\left(-\frac{1}{T}\right). \label{eq:JQGG}
\end{align}
Comparing Eqs.~\eqref{eq:JNLL}--\eqref{eq:JQLL} and Eqs.~\eqref{eq:JNGG}--\eqref{eq:JQGG}, we can express the new transport coefficients $G_{NN}$, $G_{NQ}$, $G_{QN}$, and $G_{QQ}$ in terms of $L_{NN}$, $L_{NQ}$, $L_{QN}$, and $L_{QQ}$:
\begin{align}
G_{NN}&=L_{NN}, \\
G_{NQ} &= L_{NE}-\mu L_{NN}, \\
G_{QN} &= L_{EN}-\mu L_{NN}, \\
G_{QQ}&=L_{EE}-\mu(L_{EN}+L_{NE})+\mu^2L_{NN}.
\end{align}
We can then prove the Onsager-Casimir relations for the particle and heat flows under a magnetic field $B$ as follows:
\begin{align}
G_{NN}(B) &= L_{NN}(B) = L_{NN}(-B) = G_{NN}(-B), \label{eq:GNN} \\
G_{NQ}(B) &= L_{NE}(B)-\mu L_{NN}(B)= L_{EN}(-B)-\mu L_{NN}(-B) = G_{QN}(-B), \\
G_{QQ}(B) &= L_{EE}(B)-\mu(L_{EN}(B)+L_{NE}(B))+\mu^2L_{NN}(B) \notag \\
 &= L_{EE}(-B)-\mu(L_{NE}(-B)+L_{EN}(-B))+\mu^2L_{NN}(-B) = G_{QQ}(-B), \label{eq:GQQ}
\end{align}
where we used the Onsager-Casimir relations for the particle and energy flows, Eqs.~\eqref{eq:LNN}--\eqref{eq:LEE}.
We thus conclude that if the Onsager-Casimir relations for the particle and energy flows are valid, the Onsager-Casimir relations for the particle and heat flows are also valid.
Using these relations, we can further prove that the upper limit of the efficiency Eq.~\eqref{eq:1efficiency} is the Carnot efficiency, using the Onsager-Casimir relations and the positivity of the entropy production \cite{saito2010}. 

\subsection{Thermoelectric device in mesoscopic transport systems}
\begin{figure}[ht]
\begin{center}
\includegraphics[width=13cm]{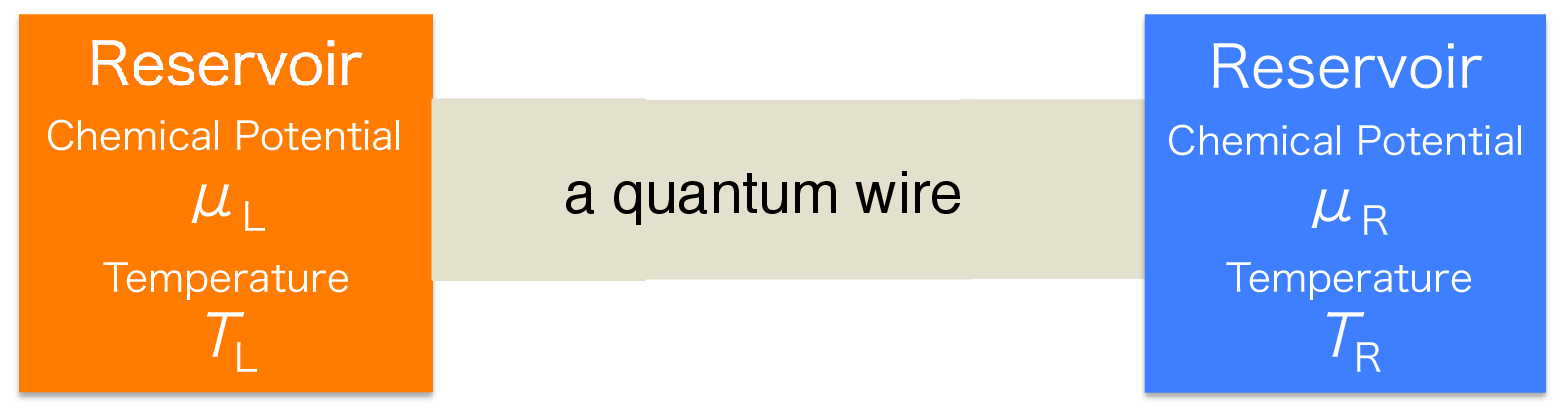}
\end{center}
\caption{The mesoscopic system which has a quasi one-dimensional wire and two reservoirs attached to it on both sides.}
\label{fig:system0}
\end{figure}
Mesoscopic transport systems are systems in which a conductor of length $L_s$  much shorter than the momentum-relaxation length $L_p$ and the phase-relaxation length $L_\phi$, is attached to reservoirs \cite{datta1995,diventra2008}. Various interesting phenomena occur because of the length scale, which is called the ballistic transport regime.  Let us consider here the simplest system shown in Fig.~\ref{fig:system0}, which consists of a quasi-one-dimensional quantum wire in the ballistic transport regime and two reservoirs attached to it on both sides. The Landauer-B\"uttiker  formula is particularly useful in this regime \cite{landauer1957,datta1995,diventra2008,kato2014}:
\begin{equation}
I = \frac{e}{h} \int_{\EG}^{\infty} d\epsilon \tau(\epsilon) (\fL(\epsilon)-\fR(\epsilon)),
\end{equation} 
where $I$ is the electric current across the system, $\EG$ the ground-state energy of the wire, $\tau(\epsilon)$ the total transmission probability of the wire, and $f_{\alpha}(\epsilon)$ the Fermi distribution function of a Fermi gas in a reservoir $\alpha=\mathrm{L,R}$; see the notation section 1.2 and Chapter 2 for details.
Such a system was theoretically considered by R.~Landauer in 1957 \cite{landauer1957}, but had not been realized experimentally until 1988 \cite{wees1988}. 
Thanks to the improvement of sub-micron technology today, more refined mesoscopic transport systems are made experimentally, which is also stimulating vigorous theoretical researches.

With the development of the research in the mesoscopic transport systems, the mesoscopic \textit{thermoelectric} device has also been considered theoretically and experimentally \cite{saito2011,brandner2013,brandner2013multi,Benenti,snyder2008}. It is expected to have a good efficiency because of its little heat leakage \cite{snyder2008}.

In such a device, nonlinear effects can occur easily. For example, an experiment \cite{matthews2014} suggests that the Onsager-Casimir relations, which are valid in the linear-response regime, are broken under a strong external field.  This implies that one can make mesoscopic thermoelectric devices in a nonlinear regime under controlled external fields such as the difference of the chemical potential and the temperature. 
However, most of the theoretical approaches are still in the linear-response regime \cite{saito2011,brandner2013,brandner2013multi,Benenti}. There are a limited number of researches in the nonlinear regime \cite{izumida2012,izumida2014}, but a general nonlinear theory for mesoscopic thermoelectric systems, which would be a counterpart of Onsager's formalism in the linear-response regime, is yet to come.

Indeed, Saito and Utsumi  \cite{saito2008} have recently found relations among nonlinear transport coefficients of the particle and energy flows, using full counting statistics \cite{nazarov2003,saito2009}.
In the linear-response regime, if the Onsager-Casimir relations for the particle and energy flows are valid, those for the particle and \textit{heat} flows are also valid as we showed above. 
There should thus be similar relations among nonlinear transport coefficients of the particle and heat flows. We have indeed found them using full counting statistics, which we explain in this thesis. 

In Chapter 2, we review the Landauer-B\"uttiker formula \cite{landauer1957,datta1995,diventra2008,kato2014}, which is essential in treating mesoscopic transport systems. Using it, we calculate the average particle flow as well as its second-order cumulant. 
In Chapter 3, we review the work by K.~Saito and Y.~Utsumi \cite{saito2008}, in which the authors obtained the relations among the transport coefficients of the particle and energy flows. 
Using the analysis of full counting statistics \cite{nazarov2003,saito2009}, we calculate the cumulant generating function of the particle and energy flows and prove its symmetry. 
The symmetry produces the relations among transport coefficients of the higher-order cumulants of the particle and energy flows. 
In Chapter 4, we properly define the heat flow in mesoscopic transport systems and derive the relations among the transport coefficients of the higher-order cumulants of the particle and \textit{heat} flows. 
Moreover, we introduce the generalized flows which are superpositions of the particle and energy flows.
We show that we can choose the appropriate affinities of the generalized flows and derive the relations among their transport coefficients when we expand the generalized flows with respect to their affinities.
We finally discuss their application to the calculation of the nonlinear Seebeck coefficient. In Chapter 5, we summarize our results and discuss possible future works.

\section{Notation}
Let us here fix the notation.
Throughout the present thesis, $\kB$ denotes the Boltzmann constant, $h$ the Planck constant and $e$ the elementary charge; for simplicity, we put $\kB=1$ throughout this thesis.

Let $f_{\alpha}(\epsilon)$ denote the Fermi distribution function of a Fermi gas in a reservoir $\alpha$: 
\begin{equation}
f_{\alpha}(\epsilon) = [1+e^{\beta_{\alpha}(\epsilon-\mu_{\alpha})}]^{-1},
\end{equation}
where $\beta_{\alpha}$ is the inverse temperature $1/T_{\alpha}$ with $T_\alpha$ the temperature and $\mu_{\alpha}$ is the chemical potential.

\chapter{Landauer-B\"uttiker Formalism}
In this chapter, we explain the Landauer-B\"uttiker formalism, calculating the first- and second-order cumulants, namely, the average and the noise of the particle flow. In Section 2.1,  We present an elementary introduction of the Landauer-B\"uttiker formalism because we would like readers to understand the formalism intuitively first. In Section 2.2, we introduce the approach using the second quantization to calculate the average as well as the noise of the particle flow. 
Note that we neglect the spin of electrons throughout this chapter. 
\section{Landauer-B\"uttiker formalism for calculation of average current}
\begin{figure}
\begin{center}
\includegraphics[width=13cm]{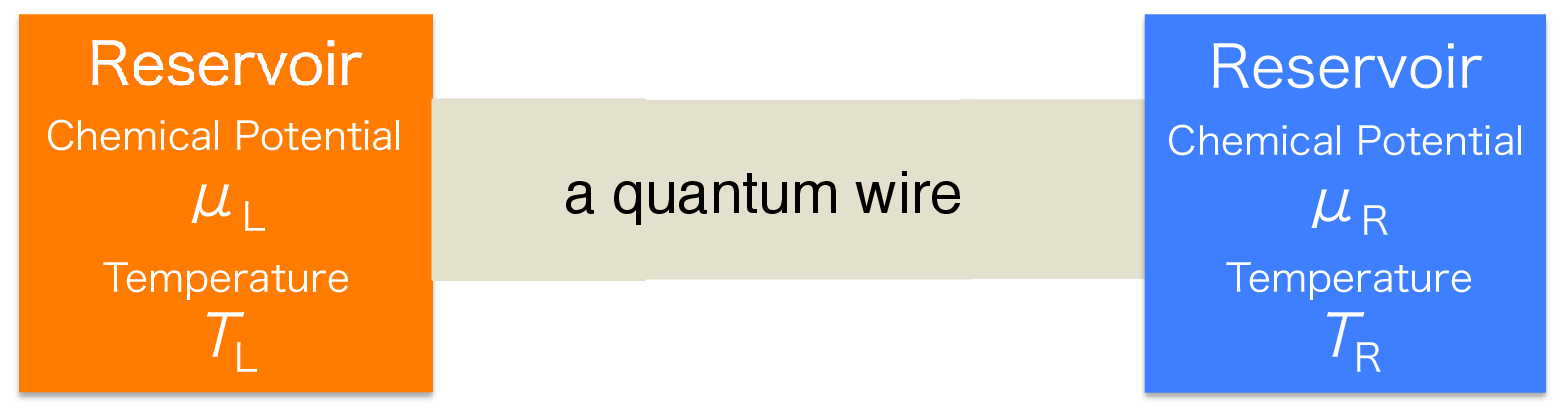}
\end{center}
\caption{The system in consideration.}
\label{fig:system}
\end{figure}

The Landauer-B\"uttiker formula is used to calculate the electric current through a quasi-one-dimensional mesoscopic conductor. 
The word `quasi-one-dimensional' indicates a conductor with the $x$-direction free while $y$- and $z$-directions confined. We specifically consider the system shown in Fig.~\ref{fig:system}, which consists of a quasi-one-dimensional quantum wire and two reservoirs attached to it on both sides. 

\begin{figure}
\begin{center}
\includegraphics[width=7cm]{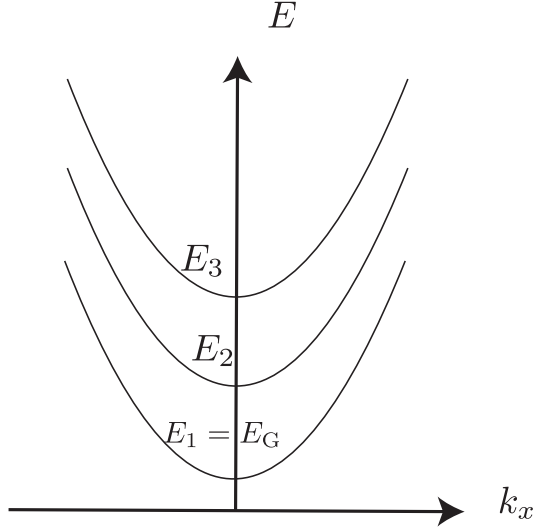}
\end{center}
\caption{Dispersion relation in the $x$ direction.}
\label{fig:dispersion}
\end{figure}

We here consider free electrons; that is, we neglect the electron-electron and electron-phonon interactions. The Schr\"odinger equation of an electron in the wire is given by 
\begin{equation}
H\Psi(x,y,z) = \left(\frac{\hbar^2\bm{k}^2}{2m} + V(y,z)\right)\Psi(x,y,z) = E\Psi(x,y,z),
\end{equation}
where $\hbar$ is the Planck constant, $\bm{k}$ is the wave-number vector, $m$ the effective mass of electrons, $V(y,z)$ the confined potential, and $E$ the energy. Let us here separate the variables of $\Psi(x,y,z)$ as follows:
\begin{equation}
\Psi(x,y,z) = \psi(x)\phi(y,z) ,
\end{equation} 
which gives the Schr\"odinger equation in each direction in the forms
\begin{align}
&\frac{\hbar^2k_x^2}{2m}\psi(x) = E_x\psi(x) , \label{eq:Sch_x} \\
&\left(\frac{\hbar^2k_y^2+\hbar^2k_z^2}{2m} + V(y,z)\right)\phi(y,z) = E_{y,z}\phi(y,z), \label{eq:Sch_yz} 
\end{align}
where each of $E_x$ and $E_{y,z}$ is the energy in the corresponding direction.  
Solving Eqs.~\eqref{eq:Sch_x} and \eqref{eq:Sch_yz}, we obtain the wave function $\Psi(x,y,z)$ as follows:
\begin{equation}
\Psi_{\pm k_x}(x,y,z) = e^{\pm ik_x x} \phi_n(y,z),
\end{equation}
where $k_x$ is the wave number in the $x$ direction given by
\begin{equation}
E_x = \frac{\hbar^2{k_x}^2}{2m}.
\end{equation} 
Note that as the electrons are confined in the $y$ and $z$ directions, they have the discrete energy $E_{y,z}=E_n$, where $n$ is the label of the level. Let us refer to the levels labeled by $n$ as `channels', through which electrons are transported in the $x$ direction. 
The total energy of an electron is 
\begin{equation}
E = E_x + E_{y,z} = \frac{\hbar^2{k_x}^2}{2m}+E_n;
\end{equation}
see Fig.~\ref{fig:dispersion}.
Throughout this thesis, we denote the energy for $k_x=0$ and $n=0$ by the ground-state energy $\EG$. 

In the Landauer-B\"uttiker formalism, the important quantity is the transmission coefficient of the wire. We therefore explain the quantum scattering problem for a while, particularly the S-matrix.
\subsection{Single-channel case}
\begin{figure}
\begin{center}
\includegraphics[width=7cm]{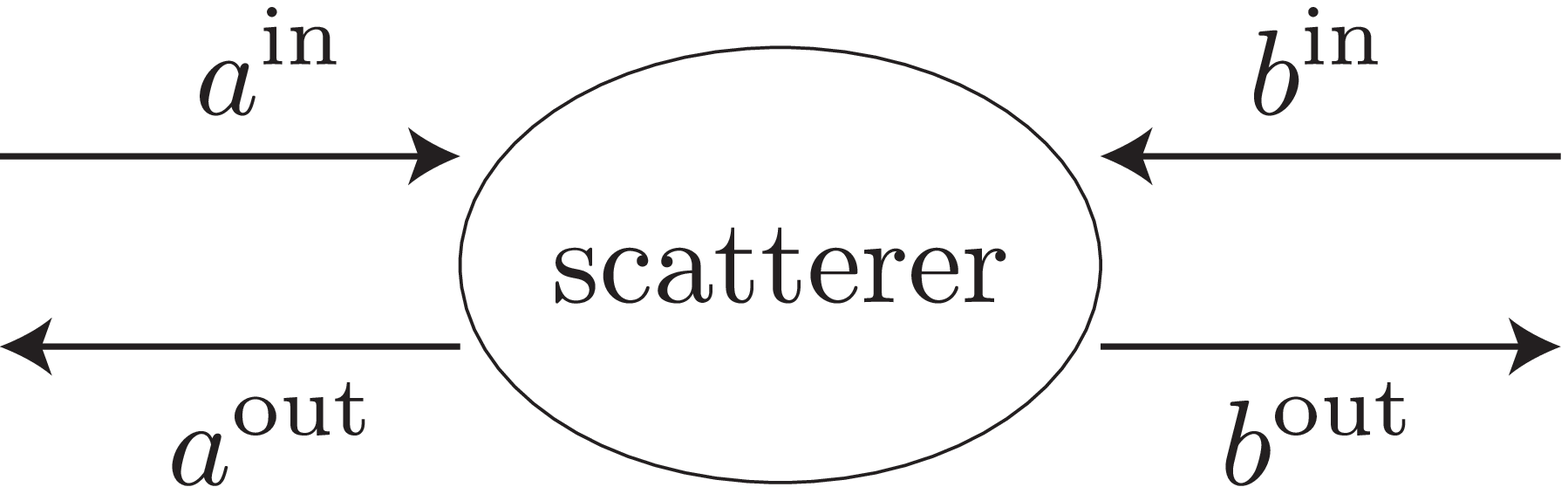}
\end{center}
\caption{The definition of the amplitudes $\ain$, $\aout$, $\bin$, and $\bout$.}
\label{fig:schannel}
\end{figure}
Let us first consider the S-matrix in the case of a single channel, in which we denote $\ain$, $\aout$, $\bin$, and $\bout$ as the amplitudes of the incoming flow from the left reservoir, the outgoing flow to the left reservoir, the incoming flow from the right reservoir, and the outgoing flow to the right reservoir, respectively (Fig.~\ref{fig:schannel}).
The relation among the amplitudes $\ain$, $\aout$, $\bin$, and $\bout$ is expressed by the S-matrix in the form
\begin{equation}
  \begin{pmatrix}
      \aout \\
      \bout \\
    \end{pmatrix}
 = S 
    \begin{pmatrix}
      \ain \\
      \bin \\
    \end{pmatrix} 
=
    \begin{pmatrix}
r & t' \\
t & r' \\
\end{pmatrix}
\begin{pmatrix}
      \ain \\
      \bin \\
    \end{pmatrix}, \label{eq:Smat}
\end{equation}
where $r$ and $r'$ are the reflection coefficients while $t$ and $t'$ are the transmission coefficients.
We here remark that $|t|^2$ and $|r|^2$ are the transmission and reflection probabilities from left to right and $|t'|^2$ and $|r'|^2$ are those from right to left.

We can prove that the S-matrix is unitary as follows. The conservation of the flux gives the conditions
\begin{equation}
|\ain|^2 +|\bin|^2 = |\aout|^2+ |\bout|^2.\label{eq:waveconsava}
\end{equation}
Using Eqs.~\eqref{eq:Smat} and \eqref{eq:waveconsava}, we have 
\begin{align}
&  |\aout|^2+ |\bout|^2 =
\begin{pmatrix}
\aout^\dagger &\bout^\dagger
\end{pmatrix}
\begin{pmatrix}
\aout \\ \bout 
\end{pmatrix}
=\begin{pmatrix}
\ain^\dagger &\bin^\dagger
\end{pmatrix}
S^\dagger S
\begin{pmatrix}
\ain \\ \bin 
\end{pmatrix}  \notag \\
=& |\ain|^2 +|\bin|^2 =
 \begin{pmatrix}
\ain^\dagger& \bin^\dagger
\end{pmatrix}
\begin{pmatrix}
\ain \\ \bin
 \end{pmatrix}
 =\begin{pmatrix}
\aout^\dagger &\bout^\dagger
\end{pmatrix}
{S^{-1}}^\dagger S^{-1}
\begin{pmatrix}
\aout \\ \bout 
\end{pmatrix}, \label{eq:Sunitary}
\end{align}
which gives $S^\dagger S = 1$ and ${S^{-1}}^\dagger S^{-1} = (SS^\dagger)^{-1} = 1$. The second condition gives $SS^\dagger = 1$. The S-matrix is therefore unitary. 

Using these unitarity conditions, we can derive the following relation among the transmission and reflection coefficients:
 \begin{align}
 SS^{\dagger} = 
 \begin{pmatrix}
 r & t' \\
 t & r'
 \end{pmatrix} 
  \begin{pmatrix}
 r ^{\dagger}& t^{\dagger} \\
 {t'}^{\dagger} & r^{\dagger}
 \end{pmatrix} 
 =
  \begin{pmatrix}
 |r|^2+|t'|^2 & r\dagg{t}+t'\dagg{r'} \\
 \dagg{r}t+r'\dagg{t'} & |r'|^2 + |t|^2
 \end{pmatrix} 
 =
 \begin{pmatrix}
 1& 0 \\
 0 & 1
 \end{pmatrix} ,
 \end{align}
which gives the condition
\begin{equation}
|r|^2+|t'|^2 = |r'|^2 + |t|^2 = 1.
\end{equation}
The relation $\dagg{S}S = 1$ also gives the condition
\begin{equation}
|r|^2+|t|^2 = |r'|^2 + |t'|^2 = 1.
\end{equation}
Using these conditions, we obtain the relations
\begin{equation}
|r|^2=|r'|^2 \quad \text{and} \quad |t|^2=|t'|^2. \label{singleT}
\end{equation}
\subsection{Multi-channel case}
\begin{figure}
\begin{center}
\includegraphics[width=7cm]{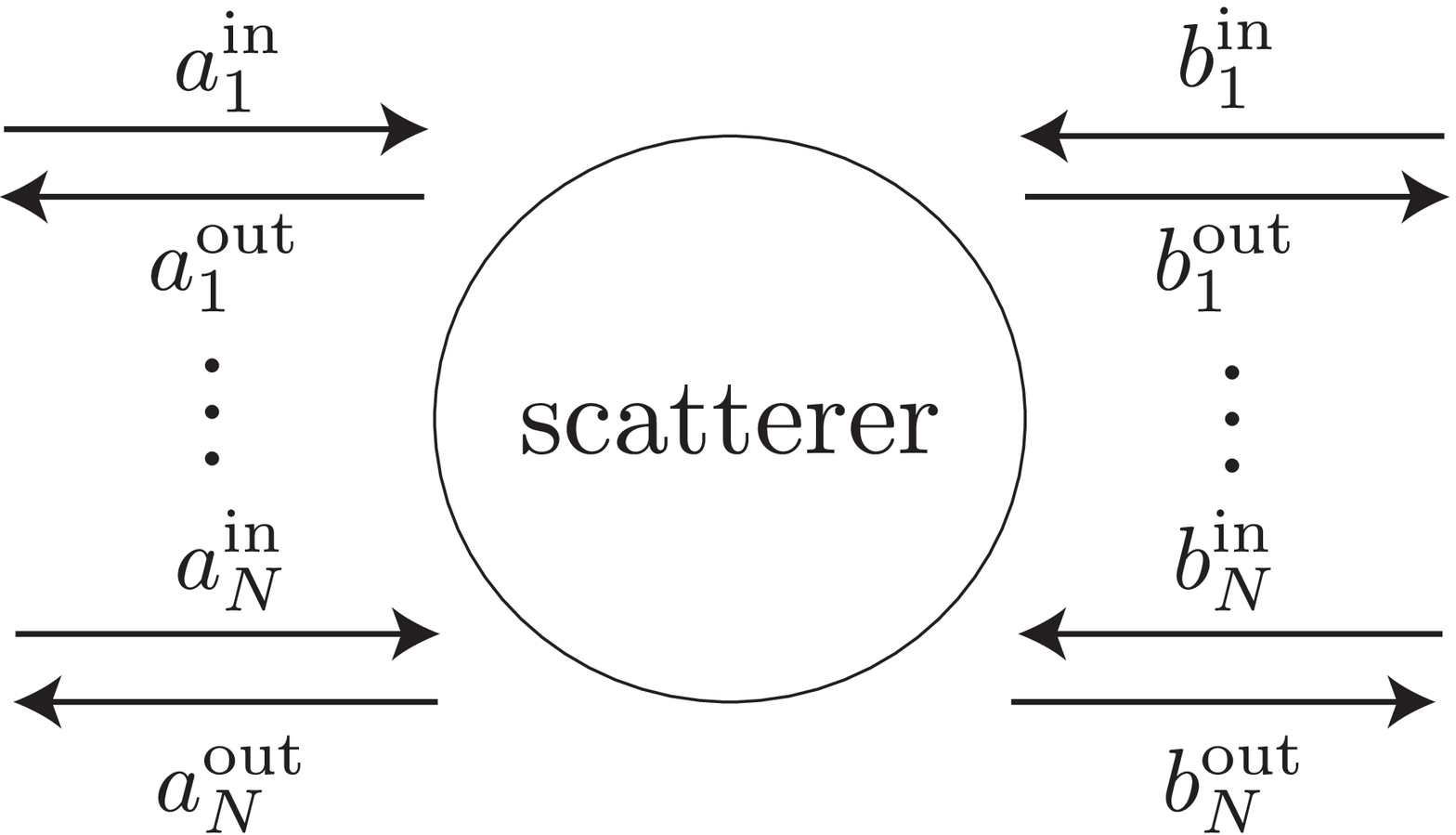}
\end{center}
\caption{The definition of the elements of $\vecain$, $\vecaout$, $\vecbin$ and $\vecbout$.}
\label{fig:mchannel}
\end{figure}
We can easily extend the single-channel case to the multi-channel case. Let $N$ denote the number of channels. Note that $N$ can be infinite in principle, but in real materials, the electron has a cut-off energy because of the energy band structure, and hence $N$ may be finite. We at first define the $N$-dimensional vectors $\vecain$, $\vecaout$, $\vecbin$ and $\vecbout$, whose  components $\anin$, $\anout$, $\bnin$, and $\bnout$ for $1\le n \le N$, respectively, denote the amplitudes of the incoming flow from the left reservoir, the outgoing flow to the left reservoir, the incoming flow from the right reservoir, and the outgoing flow to the right reservoir, each in the $n$th channel  (Fig.~\ref{fig:mchannel}). We can express the relation among these vectors with the S-matrix in the form
\begin{equation}
  \begin{pmatrix}
      \vecaout \\
      \vecbout \\
    \end{pmatrix}
 = S 
    \begin{pmatrix}
      \vecain \\
      \vecbin \\
    \end{pmatrix} 
=
    \begin{pmatrix}
r & t' \\
t & r' \\
\end{pmatrix}
\begin{pmatrix}
      \vecain \\
      \vecbin \\
    \end{pmatrix}, 
\end{equation}
where $r$ and $r'$ are $N \times N$-dimensional reflection matrices and $t$ and $t'$ are $N \times N$-dimensional transmission matrices.

Similarly to Eqs.~\eqref{eq:waveconsava}--\eqref{eq:Sunitary}, we obtain the unitarity condition $S\dagg{S} = \dagg{S}S = 1$, which gives the conditions of $r$, $r'$, $t$, and $t'$:
\begin{align}
S\dagg{S} = 1
&\Leftrightarrow \sum_{j=1}^{N}(|r_{ij}|^2+|{t'}_{ij}|^2) = \sum_{j=1}^{N}(|{r'}_{ij}|^2+|{t}_{ij}|^2) =1 , \\
\dagg{S}S = 1
&\Leftrightarrow \sum_{j=1}^{N}(|r_{ji}|^2+|{t}_{ji}|^2) = \sum_{j=1}^{N}(|{r'}_{ji}|^2+|{t'}_{ji}|^2) = 1.
\end{align}
They are followed by
\begin{align}
\sum_{j=1}^{N}(|r_{ij}|^2+|{t'}_{ij}|^2) = \sum_{j=1}^{N}(|r_{ji}|^2+|{t}_{ji}|^2), \label{eq:SS}
\end{align}
where $r_{ij}$, ${r'}_{ij}$, $t_{ij}$, ${t'}_{ij}$ are components of the reflection and the transmission matrices $r$, $r'$, $t$, $t'$, respectively. 
We here remark that  $|r_{ij}|^2$ and  $|{r'}_{ij}|^2$ are the reflection probabilities from the $j$th channel of the left and right to the $i$th channel of the left and right, respectively, while $|t_{ij}|^2$ and  $|{t'}_{ij}|^2$ are the transmission probabilities from the $j$th channel of the left and right to the $i$th channel of the right and left, respectively. 
Summing both sides of Eq.~\eqref{eq:SS} over $i$, we obtain the equality
\begin{align}
\sum_{i,j=1}^{N}(|r_{ij}|^2+|{t'}_{ij}|^2) &=  \sum_{i,j=1}^{N}(|r_{ji}|^2+|{t}_{ji}|^2) \notag \\
 &=  \sum_{i,j=1}^{N}(|r_{ij}|^2+|{t}_{ij}|^2), \label{eq:tijeq}
\end{align}   
where in the second equality we replaced $r_{ji}$ and $t_{ji}$ with $r_{ij}$ and $t_{ij}$, respectively.
We can obtain  from Eq.~\eqref{eq:tijeq} the following equality of the transmission probabilities:\begin{equation}
\sum_{i,j=1}^{N}|{t'}_{ij}|^2 = \sum_{i,j=1}^{N}|{t}_{ij}|^2, \label{eq:tijeq2}
\end{equation} 
where the left-hand side is the transmission probability from left to right and the right-hand side is that from right to left. This equality is a generalization of Eq.~\eqref{singleT} to the multi-channel case. Note that the transmission probability \eqref{eq:tijeq2} can be expressed in the form
\begin{equation}
\sum_{i,j=1}^{N}|{t}_{ij}|^2=\text{Tr}(t\dagg{t}).
\end{equation}
\subsection{Calculate the current}
Let us now calculate the electric current within the Landauer-B\"uttiker formula. We calculate it in the multi-channel case below. We make the following assumptions in order to do so:
\begin{itemize}
\item The current coming into the lead holds the Fermi distribution of the reservoir in which it originally was and relaxes in the reservoir which it goes into.
\item The current which goes from the lead into the reservoir is not reflected back into the lead.
\item The electrons in the lead are one-dimensional non-interacting Fermi particles, and therefore the current to the left and one to the right are independent of each other.    
\end{itemize}
Under these assumptions, the current which flows in the $i$th channel in the energy range [$\epsilon,\epsilon+d \epsilon$] is given by
\begin{equation}
dI_i^{\alpha}(\epsilon) = ev(\epsilon)f_{\alpha}(\epsilon)D_{\ell}(\epsilon) T_i^{\alpha}(\epsilon) d\epsilon,
\end{equation}
where $e$ is the elementary charge, $\alpha = \text{L, R}$ denotes the current from left to right and that from right to left, respectively, $v(\epsilon) = \text{d}\epsilon/\text{d}k$ is the group velocity of electrons, $D_{\ell}(\epsilon)=\text{d}k/\text{d}\epsilon = (hv(\epsilon))^{-1}$ is the density of states of one-dimensional ideal Fermi gas, and $T_{i}^\text{L}(\epsilon)$ is the transmission probability for electrons to transmit from the $i$th channel in the left lead to a channel in the right lead, while $T_{i}^\text{R}(\epsilon)$ the opposite. 

We can express $T_i^{\alpha}(\epsilon)$ in terms of the elements $t_{ij}$ of the transmission matrix; for example, we have
\begin{equation}
T_i^\text{L}({\epsilon}) = \sum_{j=1}^{N}|t_{ji}(\epsilon)|^2,
\end{equation}
because the matrix element $t_{ji}$ is the probability amplitude with which the incoming wave in the $i$th channel in the left lead transmits into the $j$th channel in the right lead as we explained below Eq.~\eqref{eq:SS}. We can similarly express $T_{i}^\text{R}(\epsilon)$ in terms of ${t'}_{ji}$ in the form
\begin{equation}
T_i^\text{R}({\epsilon}) = \sum_{j=1}^{N}|t'_{ji}(\epsilon)|^2.
\end{equation}

The total current in the energy range [$\epsilon, \epsilon+d\epsilon$] is the difference between the left-going current and the right-going current:
\begin{align}
dI(\epsilon) &= \sum_{i=1}^{N}dI_i^{\text{L}} - \sum_{i=1}^{N}dI_i^{\text{R}} \notag \\
   &= \frac{e}{h}\sum_{i,j=1}^{N}(\fL(\epsilon)|t_{ji}(\epsilon)|^2-\fR(\epsilon)|t'_{ji}(\epsilon)|^2)d\epsilon \notag \\
   &=  \frac{e}{h}\sum_{i,j=1}^{N}|t_{ij}(\epsilon)|^2(\fL(\epsilon)-\fR(\epsilon))d\epsilon,
\end{align}
where in the last equality we exchanged the dummy variables $i$ and $j$ and used Eq.~\eqref{eq:tijeq2}. 
We obtain the total current by integrating $dI$ with respect to the energy $\epsilon$:
\begin{align}
I &= \int dI(\epsilon) \notag \\
  &=  \frac{e}{h}\int_{\EG}^{\infty}d\epsilon\left[\sum_{i,j=1}^{N}|t_{ij}(\epsilon)|^2(\fL(\epsilon)-\fR(\epsilon))\right] \notag \\
 &= \frac{e}{h} \int_{\EG}^{\infty} d\epsilon \tau(\epsilon) (\fL(\epsilon)-\fR(\epsilon)), \label{eq:Landauer}
\end{align}
where $\tau(\epsilon) \equiv \sum_{i,j=1}^{N}|t_{ij}(\epsilon)|^2 = \text{Tr}(t\dagg{t})$ is the total transmission probability. Equation \eqref{eq:Landauer} is called the Landauer-B\"uttiker formula \cite{landauer1957,datta1995,diventra2008,kato2014}.
\subsection{Conductance quantization}
Let us observe the conductance quantization within the Landauer-B\"uttiker formula at zero temperature.
The Fermi distributions in the left and right reservoirs at zero temperature are respectively given by
\begin{equation}
f_{\alpha}(\epsilon) = \Theta(\epsilon-\mu_{\alpha}),
\end{equation} 
where $\Theta(\epsilon-\mu_{\alpha})$ is the step function. 
The current $I$ is given by the Landauer-B\"uttiker formula \eqref{eq:Landauer}, which in the present case reduces to
\begin{equation}
I =  \frac{e}{h} \int_{\EG}^{\infty} d\epsilon \tau(\epsilon) [\Theta(\epsilon-\muL)-\Theta(\epsilon-\muR)] . \label{eq:ecurrent}
\end{equation}
Let us calculate the conductance $G$, which is defined by
\begin{equation}
G \equiv \left. \frac{\mathrm{d}I}{\mathrm{d}V}\right|_{V=0},
\end{equation}
where $V$ is the voltage difference defined by $V =(\muL-\muR)/e$. Using Eq.~\eqref{eq:ecurrent}, we have
\begin{align}
G &=  \frac{e}{h} \int_{\EG}^{\infty} d\epsilon \tau(\epsilon)\left.\frac{\partial}{\partial V}[\Theta(\epsilon-\muL)-\Theta(\epsilon-\muR)]\right|_{V=0} \notag \\
&= \frac{e}{h}\tau(E_{\mathrm{F}}),
\end{align} 
where $E_{\mathrm{F}}=\muL=\muR$ denotes the chemical potential of the left and right reservoirs at $V=0$, namely the Fermi energy.

Let us consider the transmission coefficient $\tau(E_{\mathrm{F}})$. Because of the zero temperature, there is no electron which has an energy above $E_{\mathrm{F}}$, so that no electrons transmit from and to a channel whose bottom energy is higher than $E_{\mathrm{F}}$. This fact gives the transmission function
\begin{equation}
\tau(E_{\mathrm{F}}) = \sum_{i,j=1}^{n}|t_{ij}(E_{\mathrm{F}})|^2,
\end{equation}
where $n$ is the number of channels below $E_{\mathrm{F}}$.
Assuming that the diagonal elements $|t_{ii}(E_{\mathrm{F}})|^2$ are all equal to a constant $T$ and the other elements are equal to zero, we observe the conductance quantization:
\begin{equation}
G = \frac{e^2}{h}Tn .\label{eq:Gquant}
\end{equation} 
We can find from Eq.~\eqref{eq:Gquant} that  the conductance increases stepwise by a discrete value of $(e^2/h)T$ as the number of channels below the Fermi energy increases. This conductance quantization was indeed observed in the experiment \cite{wees1988}.
\section{Calculation of the noise with second quantization}
In this section, we calculate the shot noise within the Landauer-B\"uttiker formalism. In order to do this, we use the second-quantization approach \cite{martin2005,kato2014}. First, we derive the average current, whose result is the same as the Landauer-B\"uttiker formula \eqref{eq:Landauer}. We then derive the expression of the shot noise. The advantage of this approach is that the Pauli exclusion principle comes in naturally.

\subsection{Average current}
\begin{figure}[h]
\begin{center}
\includegraphics[width=5.5cm]{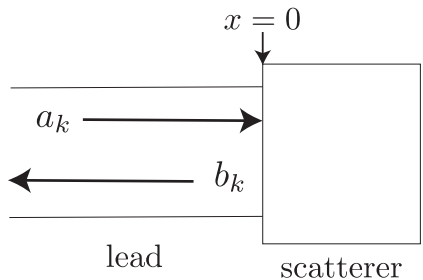}
\end{center}
\caption{The lead connected to the scatterer. We denote $\ani{a}{k}$ and $\cre{a}{k}$ the annihilation and creation operators of the right-moving electrons, while $\ani{b}{k}$ and $\cre{b}{k}$ the annihilation and creation operators of the left-moving electrons.}
\label{fig:lead_conductor}
\end{figure}
Let us first find the current operator of the system shown in Fig.~\ref{fig:lead_conductor}.
The Hamiltonian of the lead is 
\begin{equation}
H = \sum_{k}\epsilon_k\cre{c}{k}\ani{c}{k}, \label{eq:leadH}
\end{equation}
where $\ani{c}{k}$ and $\cre{c}{k}$ are the annihilation and creation operators, respectively. The energy of the electron is given by 
\begin{equation}
\epsilon_k = \frac{\hbar^2k^2}{2m}, \label{eq:dispersion}
\end{equation}
where $k$ is the wave number of the electron and $m$ the effective mass. We here remark that we may derive the Landauer-B\"uttiker formula without assuming the dispersion relation \eqref{eq:dispersion}, but for simplicity, we assume that the dispersion relation is given by Eq.{\eqref{eq:dispersion}}.
The current operator is given by
\begin{align}
I(x) &= \frac{e\hbar}{2mi}\left(\dagg{\psi}(x)\deriv{\psi}{x} - \deriv{\dagg{\psi}}{x}\psi(x) \right) \notag \\
&= \frac{e\hbar}{2imL}\sum_{k,k'}\left(ik'e^{i(k'-k)x}\cre{c}{k}\ani{c}{k'}+ike^{i(k'-k)x}\cre{c}{k}\ani{c}{k'}\right) \notag \\
 &= \frac{e\hbar}{2mL}\sum_{k,k'}(k+k')\cre{c}{k}\ani{c}{k'}e^{i(k-k')x},  \label{eq:Iope}
\end{align}
where $\psi(x)$ and $\dagg{\psi}(x)$ are the field operators given by 
\begin{align}
\psi(x) &= \frac{1}{\sqrt{L}}\sum_{k}e^{ikx}\ani{c}{k}, \\
\dagg{\psi}(x) &= \frac{1}{\sqrt{L}}\sum_{k}e^{-ikx}\cre{c}{k}.
\end{align}
Since the current is conserved, Eq.~\eqref{eq:Iope} should be independent of $x$, and hence should reduce to 
\begin{equation}
I(x) = I(0) =\frac{e\hbar}{2mL}\sum_{k,k'}(k+k')\cre{c}{k}\ani{c}{k'} . \label{eq:I(0)}
\end{equation}

In the Landauer-B\"uttiker formalism, we consider the current going to right and one going to left independently, which enables us to define the annihilation and creation operators of the right-moving electrons, $\ani{a}{k}$ and $\cre{a}{k}$, as well as the annihilation and creation operators of the left-moving electrons,  $\ani{b}{k}$ and $\cre{b}{k}$. Using these operators, we can express the Hamiltonian in the form 
\begin{equation}
H = \sum_{k>0} \epsilon_k \cre{a}{k}\ani{a}{k} +  \sum_{k<0} \epsilon_k \cre{b}{k}\ani{b}{k}.
\end{equation}
We can write down the current \eqref{eq:I(0)} in terms of the new operators as 
\begin{align}
I(x) &= I(0) = \frac{e\hbar}{2mL}\sum_{k,k'}(k+k')\cre{c}{k}\ani{c}{k'} \notag \\
 &= \frac{e\hbar}{2mL}\left[\sum_{k,k'>0}(k+k')\cre{c}{k}\ani{c}{k'} + \sum_{k,k'<0} (k+k')\cre{c}{k}\ani{c}{k'}+\sum_{k,k'\text{s.t.}kk'<0}(k+k')\cre{c}{k}\ani{c}{k'}\right] \notag \\
 &= \frac{e}{2L}\left[\sum_{k,k'>0}(v_k+v_{k'})\cre{c}{k}\ani{c}{k'} + \sum_{k,k'<0}(v_k+v_{k'})\cre{c}{k}\ani{c}{k'}+\sum_{k,k'\text{s.t.}kk'<0}(v_k+v_{k'})\cre{c}{k}\ani{c}{k'}\right] \notag \\
 &= \frac{e}{2L}\sum_{k,k'>0}(v_k+v_{k'})(\cre{c}{k}\ani{c}{k'} -\cre{c}{-k}\ani{c}{-k'})+(v_k-v_{k'})\cre{c}{k}\ani{c}{-k'}+(v_{k'}-v_{k})\cre{c}{-k}\ani{c}{k'} \notag \\
 &= \frac{e}{2L}\sum_{k,k'>0}(v_k+v_{k'})(\cre{a}{k}\ani{a}{k'} -\cre{b}{k}\ani{b}{k'})+(v_k-v_{k'})\cre{a}{k}\ani{b}{k'}+(v_{k'}-v_{k})\cre{b}{k}\ani{a}{k'} \notag \\
 &=\frac{e}{2L}\sum_{k,k'>0}(v_k+v_{k'})(\cre{a}{k}\ani{a}{k'} -\cre{b}{k}\ani{b}{k'})+(\text{cross terms}), \label{eq:IofLtoS}
\end{align}
where $v_k = \hbar k/m$ is the group velocity of the electrons. Note that the cross terms vanish when $k=k'$.

\begin{figure}
\begin{center}
\includegraphics[width=5.5cm]{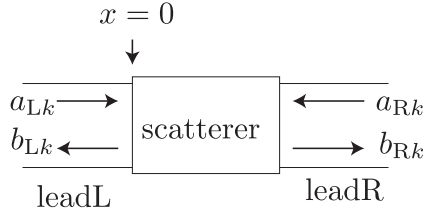}
\end{center}
\caption{Two leads attached to the both sides of the scatterer. }
\label{fig:twoleads_conductor}
\end{figure}

We next consider the system where two leads are attached to the both sides of the scatterer as shown in Fig.~\ref{fig:twoleads_conductor}. We denote the annihilation and creation operators of the incoming electrons in the lead $\alpha=\text{L,R}$ by $\dani{a}{\alpha}{k}$ and $\dcre{a}{\alpha}{k}$, respectively, while those of the outgoing electrons by $\dani{b}{\alpha}{k}$ and $\dcre{b}{\alpha}{k}$.

The S-matrix defined in Eq.~\eqref{eq:Smat} gives the relation
\begin{equation}
\begin{pmatrix}
\dlani{b}{L}{k} \\
\dlani{b}{R}{k}
\end{pmatrix}
 = S
 \begin{pmatrix}
\dlani{a}{L}{k} \\
\dlani{a}{R}{k}
\end{pmatrix},
\end{equation}
where we now use the notation
\begin{equation}
S = \begin{pmatrix}
\SLL(k) & \SLR(k) \\
\SRL(k) & \SRR(k)
\end{pmatrix}
= \begin{pmatrix}
r & t' \\
t & r'
\end{pmatrix}.
\end{equation}
The unitarity condition of the S-matrix is guaranteed by the following commutation relations:
\begin{align}
[\dani{a}{\alpha}{k}, \dcre{a}{\alpha'}{k'}] &=[\dani{b}{\alpha}{k},  \dcre{b}{\alpha'}{k'}] = \delta_{\alpha,\alpha'}\delta_{k,k'}, \\
[\dani{a}{\alpha}{k}, \dani{a}{\alpha'}{k'}] &=[\dani{b}{\alpha}{k},  \dani{b}{\alpha'}{k'}] = 0, \\
[\dcre{a}{\alpha}{k}, \dcre{a}{\alpha'}{k'}] &=[\dcre{b}{\alpha}{k},  \dcre{b}{\alpha'}{k'}] = 0. 
\end{align}
We can express the current operator $\IL(x=0)$ going from the lead L to the scatterer as Eq.~\eqref{eq:IofLtoS}:
\begin{align}
\IL(x=0) &=  \frac{e}{2L}\sum_{k,k'>0}(v_k+v_{k'})(\dlcre{a}{L}{k}\dlani{a}{L}{k'} -\dlcre{b}{L}{k}\dlani{b}{L}{k'}) +(\text{cross terms})\notag \\
  &= \frac{e}{2L}\sum_{k,k'>0}(v_k+v_{k'}) \notag \\
  &\quad \times \left[\dlcre{a}{L}{k}\dlani{a}{L}{k'}-(\conjg{\SLL(k)}\dlcre{a}{L}{k}+\conjg{\SLR(k)}\dlcre{a}{R}{k})(\SLL(k')\dlcre{a}{L}{k'}+\SLR(k')\dlcre{a}{R}{k'}) \right] \notag \\
  &\quad + (\text{cross terms}) \notag \\
  &= \frac{e}{2L}\sum_{\alpha=\text{L},\text{R}}\sum_{\beta=\text{L},\text{R}}\sum_{k,k'>0}(v_k+v_{k'})\dcre{a}{\alpha}{k}A_{\text{L}}^{\alpha\beta}(k,k')\dani{a}{\beta}{k'}+(\text{cross terms}), \label{eq:ILk}
\end{align}
where
\begin{equation}
A_{\text{L}}^{\alpha\beta}(k,k') \equiv \delta_{\text{L},\alpha}\delta_{\text{L},\beta}-\conjg{\SLa(k)}{\SLb(k')}.
\end{equation}

Let us introduce the statistical average of the creation and annihilation operators:
\begin{equation}
\expect{\dcre{a}{\alpha}{k}\dani{a}{\beta}{k'}} = \delta_{\alpha,\beta}\delta_{k,k'}f_\alpha(k), \label{eq:StatAve}
\end{equation}
which means that the distribution of incoming electrons is the Fermi distribution of the bath where they originally were. We thereby calculate the statistical average of the current as
\begin{align}
\expect{\IL} &= \frac{e}{2L}\sum_{\alpha=\text{L},\text{R}}\sum_{\beta=\text{L},\text{R}}\sum_{k,k'>0}(v_k+v_{k'})A_{\text{L}}^{\alpha\beta}(k,k')\expect{\dcre{a}{\alpha}{k}\dani{a}{\beta}{k'}}+(\text{cross terms}) \notag \\
 &= \frac{e}{2L}\sum_{\alpha=\text{L},\text{R}}\sum_{k>0}2v_kA_{\text{L}}^{\alpha\alpha}(k,k)f_\alpha(k) \notag \\
 &= \frac{e}{h}\int_{\EG}^{\infty}d\epsilon \sum_{\alpha=\text{L},\text{R}}A_{\text{L}}^{\alpha\alpha}(\epsilon,\epsilon)f_\alpha(\epsilon), \label{eq:exsl}
\end{align}
where in the second equality the cross terms vanish because they survive only when $k\neq k'$ but the statistical average insists $k=k'$.
In the last equality we replaced the summation $\sum_{k}$ with the integral with respect to energy, assuming that the value of $L$ is large enough for us to do the transformation as follows:
\begin{equation}
\frac{1}{L} \sum_{k} \to \int\frac{dk}{2\pi} \to \int d\epsilon\Dl(\epsilon), \label{eq:k_to_e}
\end{equation} 
where 
\begin{equation}
\Dl(\epsilon) = \frac{1}{2\pi}\frac{\text{d}k}{\text{d}E} = \frac{1}{2\pi \hbar v_k} = \frac{m}{2\pi\hbar^2k}
\end{equation}
 is the density of states of the one-dimensional ideal Fermi gas.

Using the relations between the elements of the S-matrix, we have
\begin{align}
A_{\text{L}}^{\text{L}\text{L}}(\epsilon,\epsilon) &= 1-\conjg{\SLL(\epsilon)}\SLL(\epsilon) = 1-|r|^2 = |t|^2 \equiv\tau(\epsilon), \label{eq:ALLL} \\ 
A_{\text{L}}^{\text{R}\text{R}}(\epsilon,\epsilon) &= -\conjg{\SLR(\epsilon)}\SLR(\epsilon) =-|t'|^2=-|t|^2 \equiv -\tau(\epsilon), \label{eq:ALRR}
\end{align}
which reduces \eqref{eq:exsl} to the Landauer-B\"uttiker formula
\begin{equation}
\expect{\IL} = \frac{e}{h}\int_{\EG}^{\infty}d\epsilon \tau(\epsilon)(\fL(\epsilon)-\fR(\epsilon)).
\end{equation}

\subsection{Shot noise}
The current noise contains much important information of the system in question. The power spectrum of the current noise is defined by 
\begin{equation}
S(\omega) = \lim_{T\to\infty}\frac{2}{T}|\Delta I(\omega)|^2 = \lim_{T\to\infty}\frac{2}{T}\int_{-\frac{T}{2}}^{\frac{T}{2}}dt\int_{-\frac{T}{2}}^{\frac{T}{2}}dt'\expect{\Delta I(t) \Delta I(t')}e^{i\omega(t-t')}, \label{eq:Somega} 
\end{equation}
where $\omega$ is the frequency, $T$ is the time to observe the current, $I(t)$ is the electric current, $\Delta I(t) \equiv \expect{I(t)}-I(t)$ is the fluctuation of the electric current from its average value, and $I(\omega) = \int_{-T/2}^{T/2}dtI(t)e^{i\omega t}$ is the Fourier transform of the electric current $I(t)$. 
To calculate the shot noise, let us define the correlation function of the electric current as follows:
\begin{equation}
C(t,t') = \expect{\Delta\IL(t)\Delta\IL(t')} = \expect{\IL(t)\IL(t')}-\expect{\IL(t)}\expect{\IL(t')},
\end{equation}
where $\Delta \IL(t) = \IL(t)-\expect{\IL(t)}$ is the fluctuation operator of the current. Using this function, we can express the power spectrum of the current noise, Eq.~\eqref{eq:Somega}, as follows:
\begin{equation}
S(\omega) = \lim_{T\to\infty}\frac{2}{T}\int_{-\frac{T}{2}}^{\frac{T}{2}}dt\int_{-\frac{T}{2}}^{\frac{T}{2}}dt'C(t,t')e^{i\omega(t-t')},
\end{equation}
which is of the same form as the classical noise power. Note, however, that the current operators $\IL(t)$ and $\IL(t')$ of different time do not commute with each other in the quantum case. 
When the Hamiltonian of the system does not depend on time, the system has the time-translational symmetry, and hence we can express the correlation function as $C(t,t') = C(t-t')$. Moreover, we assume that the correlation function $C(t-t')$ tends to zero when the time difference $|t-t'|$ goes to infinity. Based on  these considerations, we can express the noise power $S(\omega)$ in the form: 
\begin{align}
S(\omega)  &= \lim_{T\to\infty}\frac{2}{T}\int_{-\frac{T}{2}}^{\frac{T}{2}}dt\int_{-\frac{T}{2}}^{\frac{T}{2}}dt'C(t-t')e^{i\omega(t-t')} \notag \\
 &=  \lim_{T\to\infty}\frac{2}{T}\int_{-\frac{T}{2}}^{\frac{T}{2}}dt'\int_{-T}^{T}d\Delta t C(\Delta t)e^{i\omega\Delta t} \notag \\
 &\approx  \lim_{T\to\infty}\frac{2}{T}\int_{-\frac{T}{2}}^{\frac{T}{2}}dt'\int_{-\infty}^{+\infty}d\Delta t C(\Delta t)e^{i\omega\Delta t} \notag \\
 &= 2\int_{-\infty}^{\infty}dt C(t) e^{i\omega t},
\end{align}

In general, $S(\omega)$ contains many components of different $\omega$. We, however, only calculate the zero-frequency component of $S(\omega)$ for simplicity:
\begin{equation}
S(0) = 2\int_{-\infty}^{\infty}dt  \expect{\Delta\IL(t)\Delta\IL(0)} = 2\int_{-\infty}^{\infty}dt\left(\expect{\IL(t)\IL(0)}-\expect{\IL(t)}\expect{\IL(0)}\right) . \label{eq:S(0)}
\end{equation}
In order to calculate $S(0)$, we need the time evolution of the current operator $\IL(t)$. We thus use the time evolution of the creation and annihilation operators:
\begin{align}
\dani{a}{\alpha}{k}(t) &= e^{-i\epsilon_kt/\hbar}\dani{c}{\alpha}{k}, \\
\dcre{a}{\alpha}{k}(t) &= e^{i\epsilon_kt/\hbar}\dcre{c}{\alpha}{k},
\end{align}
which gives us the expression of the current operator at time $t$:
\begin{equation}
\IL(t) =  \frac{e}{2L}\sum_{\alpha=\text{L},\text{R}}\sum_{\beta=\text{L},\text{R}}\sum_{k,k'>0}(v_k+v_{k'})\dcre{a}{\alpha}{k}A_{\text{L}}^{\alpha\beta}(k,k')\dani{a}{\beta}{k'}e^{i(\epsilon_k-\epsilon_{k'})t/\hbar},
\end{equation}
where we ignore the cross terms because they will vanish when we take the statistical average below.
Substituting Eq.~\eqref{eq:ILk} and its time evolution $\IL(t)$ into Eq.~\eqref{eq:S(0)}, we can express $S(0)$ in the form:
\begin{align}
S(0) &= 2\left(\frac{e}{2L}\right)^2 \int_{-\infty}^{\infty}dt \sum_{k,k',k'',k'''>0}\sum_{\alpha,\beta,\alpha'\beta'}(v_k+v_{k'})(v_{k''}+v_{k'''})A_{\text{L}}^{\alpha\beta}(k,k')A_{\text{L}}^{\alpha'\beta'}(k'',k''') \notag \\
  & \quad \times \left[ \expect{\dcre{a}{\alpha}{k}\dani{a}{\beta}{k'}\dcre{a}{\alpha'}{k''}\dani{a}{\beta'}{k'''}}-\expect{\dcre{a}{\alpha}{k}\dani{a}{\beta}{k'}}\expect{\dcre{a}{\alpha'}{k''}\dani{a}{\beta'}{k'''}}\right]e^{i(\epsilon_k-\epsilon_k')t/\hbar}.
\end{align}

In order to calculate the statistical average of the creation and annihilation operators, we use Wick's theorem \cite{fetter2003}:
\begin{equation}
\expect{ABCD} = \expect{AB}\expect{CD}-\expect{AC}\expect{BD}+\expect{AD}\expect{BC},
\end{equation}
where $A$, $B$, $C$, and $D$ are arbitrary Fermion operators.  Substituting $\dcre{a}{\alpha}{k}$ for $A$, $\dani{a}{\beta}{k'}$ for $B$, $\dcre{a}{\alpha'}{k''}$ for $C$, and $\dani{a}{\beta'}{k'''}$ for $D$, we obtain 
\begin{align}
\expect{\dcre{a}{\alpha}{k}\dani{a}{\beta}{k'}\dcre{a}{\alpha'}{k''}\dani{a}{\beta'}{k'''}} 
  &= \expect{\dcre{a}{\alpha}{k}\dani{a}{\beta}{k'}}\expect{\dcre{a}{\alpha'}{k''}\dani{a}{\beta'}{k'''}}-\expect{\dcre{a}{\alpha}{k}\dcre{a}{\alpha'}{k''}}\expect{\dani{a}{\beta}{k'}\dani{a}{\beta'}{k'''}} \notag \\
  & \quad +\expect{\dcre{a}{\alpha}{k}\dani{a}{\beta'}{k'''}}\expect{\dani{a}{\beta}{k'}\dcre{a}{\alpha'}{k''}} \notag \\
  &= \expect{\dcre{a}{\alpha}{k}\dani{a}{\beta}{k'}}\expect{\dcre{a}{\alpha'}{k''}\dani{a}{\beta'}{k'''}} +\expect{\dcre{a}{\alpha}{k}\dani{a}{\beta'}{k'''}}\expect{\dani{a}{\beta}{k'}\dcre{a}{\alpha'}{k''}}. \label{eq:Wick}
\end{align}
In the second equality, we used the equation
\begin{equation}
\expect{\dcre{a}{\alpha}{k}\dcre{a}{\alpha'}{k''}}\expect{\dani{a}{\beta}{k'}\dani{a}{\beta'}{k'''}} = 0,
\end{equation}
because the operators $\dcre{a}{\alpha}{k}\dcre{a}{\alpha'}{k''}$ and $\dani{a}{\beta}{k'}\dani{a}{\beta'}{k'''}$ do not conserve the number of particles, so that the statistical averages of these operators become zero.
Using Eqs.~\eqref{eq:StatAve} and \eqref{eq:Wick}, we obtain the following relation:
\begin{align}
 \expect{\dcre{a}{\alpha}{k}\dani{a}{\beta}{k'}\dcre{a}{\alpha'}{k''}\dani{a}{\beta'}{k'''}}-\expect{\dcre{a}{\alpha}{k}\dani{a}{\beta}{k'}}\expect{\dcre{a}{\alpha'}{k''}\dani{a}{\beta'}{k'''}} &= \expect{\dcre{a}{\alpha}{k}\dani{a}{\beta'}{k'''}}\expect{\dcre{a}{\beta}{k'}\dani{a}{\alpha}{k''}} \notag \\
 &= \delta_{\alpha,\beta'}\delta_{k,k'''}\delta_{\beta,\alpha'}\delta_{k',k''}f_{\alpha}(k)(1-f_{\beta}(k)). \label{eq:StatAve2}
\end{align}
Using Eq.~\eqref{eq:StatAve2} and the transformation \eqref{eq:k_to_e}, we finally obtain the expression of the shot noise as follows:
\begin{align}
S(0) &=2\left(\frac{e}{2L}\right)^2 \int_{-\infty}^{\infty}dt \sum_{k,k',k'',k'''>0}\sum_{\alpha,\beta,\alpha'\beta'}(v_k+v_{k'})(v_{k''}+v_{k'''})A_{\text{L}}^{\alpha\beta}(k,k')A_{\text{L}}^{\alpha'\beta'}(k'',k''') \notag \\
 & \quad \times \left[ \expect{\dcre{a}{\alpha}{k}\dani{a}{\beta}{k'}\dcre{a}{\alpha'}{k''}\dani{a}{\beta'}{k'''}}-\expect{\dcre{a}{\alpha}{k}\dani{a}{\beta}{k'}}\expect{\dcre{a}{\alpha'}{k''}\dani{a}{\beta'}{k'''}}\right]e^{i(\epsilon_k-\epsilon_k')t/\hbar} \notag \\
 &= 2\left(\frac{e}{2L}\right)^2 \int_{-\infty}^{\infty}dt \sum_{k,k',k'',k'''>0}\sum_{\alpha,\beta,\alpha'\beta'}(v_k+v_{k'})(v_{k''}+v_{k'''})A_{\text{L}}^{\alpha\beta}(k,k')A_{\text{L}}^{\alpha'\beta'}(k'',k''') \notag \\
 & \quad \times \left[\delta_{\alpha,\beta'}\delta_{k,k'''}\delta_{\beta,\alpha'}\delta_{k',k''}f_{\alpha}(k)(1-f_{\beta}(k))\right]e^{i(\epsilon_k-\epsilon_k')t/\hbar} \notag \\
 &= 2\left(\frac{e}{2L}\right)^2 \int_{-\infty}^{\infty}dt \sum_{k,k'>0}\sum_{\alpha,\beta,}(v_k+v_{k'})^2A_{\text{L}}^{\alpha\beta}(k,k')A_{\text{L}}^{\beta\alpha}(k',k)f_{\alpha}(k)(1-f_{\beta}(k))e^{i(\epsilon_k-\epsilon_k')t/\hbar} \notag \\
 &=  \frac{e^2}{2h^2}\int_{\EG}^{\infty}d\epsilon\Dl(\epsilon)\int_{\EG}^{\infty}d\epsilon' \Dl(\epsilon')\int_{-\infty}^{\infty}dt \notag \\
 & \quad \times \sum_{\alpha,\beta}(v_k+v_{k'})^2A_{\text{L}}^{\alpha\beta}(\epsilon,\epsilon')A_{\text{L}}^{\beta\alpha}(\epsilon',\epsilon)f_{\alpha}(\epsilon)(1-f_{\beta}(\epsilon))e^{i(\epsilon-\epsilon')t/\hbar} \notag \\
 &= \frac{2e^2}{h}\int_{\EG}^{\infty}d\epsilon\sum_{\alpha,\beta}A_{\text{L}}^{\alpha\beta}(\epsilon,\epsilon)A_{\text{L}}^{\beta\alpha}(\epsilon,\epsilon)f_{\alpha}(\epsilon)(1-f_{\beta}(\epsilon)), \label{eq:noiseresult}
\end{align}
where we used the relation
\begin{equation}
\int_{-\infty}^{\infty}dt e^{i(\epsilon-\epsilon')t/\hbar} = 2\pi\hbar\delta(\epsilon-\epsilon')
\end{equation}
in the last equality.

In order to express the noise more simply, we use Eqs.~\eqref{eq:ALLL} and \eqref{eq:ALRR} as well as the relations
\begin{align}
A_{\text{L}}^{\text{L}\text{R}}(\epsilon,\epsilon) &= -\conjg{\SLL(\epsilon)}\SLR(\epsilon) = -\conjg{r}t', \\
A_{\text{L}}^{\text{R}\text{L}}(\epsilon,\epsilon) &=-\conjg{\SLR(\epsilon)}\SLL(\epsilon) = -\conjg{t'}r,
\end{align}
which give the relations
\begin{equation}
A_{\text{L}}^{\text{L}\text{R}}(\epsilon,\epsilon)A_{\text{L}}^{\text{R}\text{L}}(\epsilon,\epsilon)=A_{\text{L}}^{\text{R}\text{L}}(\epsilon,\epsilon)A_{\text{L}}^{\text{L}\text{R}}(\epsilon,\epsilon) = |r|^2|t'|^2 = |r|^2|t'|^2 = \tau(\epsilon)(1-\tau(\epsilon)).
\end{equation}
We then calculate the integrand of Eq.~\eqref{eq:noiseresult} as
\begin{align}
& \sum_{\alpha,\beta}A_{\text{L}}^{\alpha\beta}(\epsilon,\epsilon)A_{\text{L}}^{\beta\alpha}(\epsilon,\epsilon)f_{\alpha}(\epsilon)(1-f_{\beta}(\epsilon)) \notag \\
& \quad= A_{\text{L}}^{\text{L}\text{L}}(\epsilon,\epsilon)A_{\text{L}}^{\text{L}\text{L}}(\epsilon,\epsilon)\fL(\epsilon)(1-\fL(\epsilon)) + A_{\text{L}}^{\text{L}\text{R}}(\epsilon,\epsilon)A_{\text{L}}^{\text{R}\text{L}}(\epsilon,\epsilon)\fL(\epsilon)(1-\fR(\epsilon)) \notag \\
 &\qquad +A_{\text{L}}^{\text{R}\text{L}}(\epsilon,\epsilon)A_{\text{L}}^{\text{L}\text{R}}(\epsilon,\epsilon)\fR(\epsilon)(1-\fL(\epsilon))+A_{\text{L}}^{\text{R}\text{R}}(\epsilon,\epsilon)A_{\text{L}}^{\text{R}\text{R}}(\epsilon,\epsilon)\fR(\epsilon)(1-\fR(\epsilon)) \notag \\
  &\quad = \tau(\epsilon)^2[\fL(\epsilon)(1-\fL(\epsilon))+\fR(\epsilon)(1-\fR(\epsilon))] \notag \\
   & \qquad + \tau(\epsilon)(1-\tau(\epsilon))[\fL(\epsilon)(1-\fR(\epsilon))+\fR(\epsilon)(1-\fL(\epsilon))] \notag \\ 
   &\quad = \tau(\epsilon)(\fL(\epsilon)+\fR(\epsilon)-2\fL(\epsilon)\fR(\epsilon))- \tau(\epsilon)^2(\fL(\epsilon)-\fR(\epsilon))^2,
\end{align}
which gives the final result of the expression of the shot noise:
\begin{equation} 
S(0) =  \frac{4e^2}{h}\int_{\EG}^{\infty}d\epsilon \left[ \tau(\epsilon)(\fL(\epsilon)+\fR(\epsilon)-2\fL(\epsilon)\fR(\epsilon))- \tau(\epsilon)^2(\fL(\epsilon)-\fR(\epsilon))^2\right].
\end{equation}

\chapter{Full Counting Statistics}
In the previous chapter, we considered the first- and second-order cumulants of the particle flow.
In this chapter, we consider the higher-order cumulants of the particle and energy flows. 
We review derivation \cite{saito2008} of the generalized Onsager relations among their transport coefficients using full counting statistics \cite{nazarov2003,saito2009}.
In Section 3.1, we introduce the full counting statistics \cite{nazarov2003,saito2009} briefly. 
In Section 3.2, we review Ref.~\cite{saito2008}. 
We calculate the cumulant generating function of the particle and energy flows and prove its symmetry. 
This symmetry gives the generalized Onsager relations for the particle and energy flows.
\section{What is counting statistics?}
Counting statistics is an analysis method in which one counts a physical quantity in time $s$ and examines its statistics. We would like to apply counting statistics to the transport phenomena driven by thermodynamic power (affinity), such as electrical conduction and heat conduction. We here explain a concept of full counting statistics \cite{nazarov2003,saito2009} using the example of electrical conduction.

Consider two reservoirs which have different chemical potentials $\muL$ and $\muR$. The difference of chemical potentials, $\muL-\muR$, causes an electric current. We here assume $\muL > \muR$ so that electrons may flow from right to left. We measure the number of the electrons $Q$ for the time $s$. After we repeat the measurement many times, we obtain the probability distribution $P(Q)$ of the number of electrons $Q$. Our aim is to know the probability distribution $P(Q)$ for large enough $s$ because we would like to know the non-equilibrium steady state, a long-time behavior of the system. 

We then explain how we obtain information of the cumulant from $P(Q)$. Let us define the characteristic function $Z(\chi)$, the Fourier transform of $P(Q)$:
\begin{equation}
Z(\chi) \equiv \sum_{Q} P(Q)e^{i\chi Q}, \label{eq:ichiQ}
\end{equation}
where $\chi$ is a variable called `counting field'. We can compute the expectation value of $Q^n$ from $Z(\chi)$ as follows:
\begin{align}
\left. \frac{\text{d}^nZ(\chi)}{\text{d}(i\chi)^n}\right|_{\chi=0} &=\left. \frac{\text{d}^n}{\text{d}(i\chi)^n}\sum_{Q} P(Q)e^{i\chi Q}\right|_{\chi=0} \notag \\
 &= \sum_{Q} P(Q)Q^n \notag \\
 &= \expect{Q^n}.
\end{align} 
Let us define the cumulant $\cumulant{Q^n}$ of $Q$ as follows:
\begin{equation}
\cumulant{Q^n} =\left. \frac{\text{d}^n\log Z(\chi)}{\text{d}(i\chi)^n} \right|_{\chi=0}.
\end{equation}
For example, $\cumulant{Q} = \expect{Q}$ is the expectation value of $Q$ and $\cumulant{Q^2} = \expect{Q^2} -\expect{Q}^2$ is the variance of $Q$.

Let us also define the cumulant generating function for the electric current:
\begin{equation}
F(\chi)\equiv \lim_{s\to\infty}\frac{1}{s}\log{Z}. 
\end{equation}
We can, for example, calculate the cumulant of the electric current $I$ as fallows:
\begin{equation}
\cumulant{I} =\left. \frac{\partial F(\chi)}{\partial(i\chi)}\right|_{\chi=0} = \lim_{s\to\infty}\frac{\expect{Q}}{s},
\end{equation}
which is the expectation value of the electric current,
and
\begin{equation}
\cumulant{I^2} = \left.\frac{\partial^2F(\chi)}{\partial(i\chi)^2}\right|_{\chi=0} = \lim_{s\to\infty}\frac{\expect{Q^2}-\expect{Q}^2}{s},
\end{equation}
which is the noise of the electric current.
We thus obtain higher-order cumulants from $F(\chi)$. 

Using this function $F(\chi)$, we can obtain an asymptotic expression of $P(Q)$ as follows:
\begin{align}
P(Q) &= \frac{1}{2\pi}\int_{-\infty}^{\infty}d\chi Z(\chi)e^{-i\chi Q} \notag \\
        &= \frac{1}{2\pi}\int_{-\infty}^{\infty}d\chi e^{\log Z(\chi) -i\chi Q} \notag \\
        &=  \lim_{s\to\infty} \frac{1}{2\pi}\int_{-\infty}^{\infty}d\chi e^{s h(\chi)},
\end{align}
where $h(\chi) \equiv F(\chi) - i\chi q$ and $q = Q/s$. We can evaluate the integral by the saddle-point method. Let $\chi^{\star}$ denote the value of $\chi$ which makes $h(\chi)$ maximum:
\begin{equation}
\left.\deriv{h(\chi)}{\chi}\right|_{\chi=\chi^\star} = 0, \quad \text{or} \quad \left.\deriv{F(\chi)}{\chi}\right|_{\chi=\chi^\star} = i\frac{Q}{s}.
\end{equation}
Expanding $h(\chi)$ around $\chi=\chi^\star$ to the second order, we arrive at an asymptotic expression of $P(Q)$ as follows:
\begin{align}
P(Q) &= \lim_{s\to\infty} \frac{1}{2\pi}\int_{-\infty}^{\infty}d\chi\exp\left[s\left(h(\chi^\star)+\frac{1}{2}(\chi-\chi^\star)^2\left. \frac{\text{d}^2h(\chi)}{\text{d}\chi^2}\right|_{\chi=\chi^\star}\right)\right] \notag \\
 & =\lim_{s\to\infty} e^{s h(\chi^\star)} = \lim_{s\to\infty}e^{s F(\chi^\star)-i\chi^\star Q}.
\end{align}
\section{Review of ref.~\cite{saito2008}}
\subsection{Setup and model Hamiltonian}
\begin{figure}[h]
\begin{center}
\includegraphics[width=7.5cm]{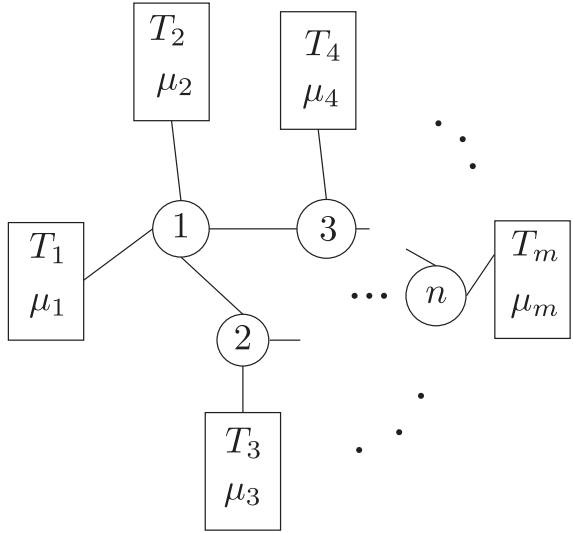}
\end{center}
\caption{The system in consideration.}
\label{fig:4system}
\end{figure}
We consider the model in which $n$ quantum dots are connected to $m$ reservoirs as shown in Fig.~\ref{fig:4system}. 
The Hamiltonian of the model is given by 
\begin{equation}
H = \sum_{r=1}^m H_r + H_d + H_{\text{int}} + H_T,
\end{equation}
where $H_r$ denotes the Hamiltonian of the $r$th reservoir, $H_d$ the Hamiltonian of the quantum dots, $H_{\text{int}}$ the Coulomb interaction between electrons on each dot as well as between the dots, and $H_T$ the tunneling Hamiltonian:
\begin{equation}
H_r = \sum_{k\sigma}\epsilon_{rk}a^{\dagger}_{rk\sigma}a_{rk\sigma},
\end{equation}
where $k$ is the wave number, $\sigma$ denotes spin, $\epsilon_{rk}$ is the energy spectrum of the $r$th reservoir, $a^{\dagger}_{rk\sigma}$ the creation operator of the reservoir, and  $a_{rk\sigma}$  the annihilation operator of the reservoir;
\begin{equation}
H_d = \sum_{i\sigma}\epsilon_{i}d^{\dagger}_{i\sigma}d_{i\sigma} +\sum_{ij\sigma}t_{ij}d^{\dagger}_{i\sigma}d_{j\sigma},
\end{equation}
where $\epsilon_{i}$ is the energy level of the $i$th dot, $t_{ij}$ is the hopping matrix between the dots, $d^{\dagger}_{i\sigma}$ the creation operator of the $i$th dot, and $d_{i\sigma}$ the annihilation operator of the $i$th dot;
\begin{equation}
H_{\text{int}} = \frac{1}{2}\sum_{ij\sigma\sigma'} U_{i\sigma j\sigma'}d^{\dagger}_{i\sigma}d^{\dagger}_{j\sigma'}d_{j\sigma'}d_{i\sigma},
\end{equation}
where $U_{i\sigma j\sigma'}$ is the strength of Coulomb interaction on each dot and between the dots;
\begin{equation}
H_T = \sum_{rki\sigma}t_{rki}d^{\dagger}_{i\sigma}a_{rk\sigma} + \text{H.c.},
\end{equation}
where $t_{rki}$ is the tunneling matrix between the $r$th reservoir and the $i$th dot.

We later consider the situation where we measure the system at time $t=-s/2$ and $t=s/2$. We assume that the initial density matrix $\rho_0$ at the time $t=-s/2$  is 
\begin{align}
\rho_0 &= \rho_d\otimes \rho_0' \notag \\
 &= \rho_d \otimes \frac{\exp[-\beta_r(H_r-\mu_rN_r)]}{\text{Tr}\{\exp[-\beta_r(H_r-\mu_rN_r)]\}}, \label{eq:rho_0'}
\end{align}
where $\rho_d$ and $\rho_0'$ are the density matrices of the quantum dots and the reservoirs, respectively, $r$  the index denoting the reservoirs and $d$ the quantum dots, $\beta_r$ the inverse temperature of the $r$th reservoir, $\mu_r$ its chemical potential, and $N_r$ its number operator:
$N_r = \sum_{k\sigma}a^{\dagger}_{rk\sigma}a_{rk\sigma}$. We assume that there had been no interaction between the reservoirs and the quantum dots until the initial time $t=-s/2$ and hence their density operators are commutative then.

We here assume that $\rho_d = 1_{d}/2^n$, where $1_d$ is the identity matrix. Because we expect that the stationary state in the long-time limit is independent of the initial state of the dots, we can arbitrarily choose the initial state of the dots, and thus took  $\rho_d = 1_{d}/2^n$ for our convenience. The assumption becomes useful in proving the symmetry of the characteristic function Eq.~\eqref{eq:Zsymm1}, whose details are shown in Appendix A.

We then define the particle current operator and the energy current operator with the Heisenberg equation:
\begin{align}
J_{Nr} &= \dot{N_r} = i[N_r,H_T] = -\sum_{ik}it_{rki}d^{\dagger}_{i\sigma}a_{rk\sigma} + \text{H.c.}, \\
J_{Er} &= \dot{H_r} = i[H_r,H_T] = -\sum_{ik}i\epsilon_{rk}t_{rki}d^{\dagger}_{i\sigma}a_{rk\sigma} + \text{H.c.}.
\end{align}
Let us define the charge $q_{Nr}$ and the energy $q_{Er}$ of the $r$th reservoir during the measuring time $s$ as follows:
\begin{align}
q_{Nr} &= \int_{-s/2}^{s/2}dtJ_{Nr}(t), \\
q_{Er} &=  \int_{-s/2}^{s/2}dtJ_{Er}(t).
\end{align}
\subsection{Protocol of the measurement}
In order to obtain the expressions of the characteristic function $Z(\chi)$ and the cumulant generating function $F(\chi)$, we need to know the probability distribution $P(Q)$ as we discussed in Sec. 3.1. We can obtain the expression of $P(Q)$, considering the following protocol of measurement \cite{kurchan2001}.
\begin{enumerate}
\item At the initial time $t=-s/2$, we carry out a projection measurement on a reservoir $r$, and thereby obtain the result $Q_{Nr,\nu}^{\rm{init}}$ and the energy $Q_{Er,\nu}^{\mathrm{init}}$ of the particles in it, which collapses the wave function of each reservoir into an eigenfunction $\ket{\psi_{r,\nu}^{\rm{init}}}$ of the particle and energy operators:
\begin{align}
N_r\ket{\psi_{r,\nu}^{\rm{init}}} &= Q_{Nr,\nu}^{\rm{init}}\ket{\psi_{r,\nu}^{\rm{init}}}, \\
H_r\ket{\psi_{r,\nu}^{\rm{init}}} &= Q_{Er,\nu}^{\rm{init}}\ket{\psi_{r,\nu}^{\rm{init}}}.
\end{align}
Note that the set of kets $\{\ket{\psi_{r,\nu}^{\rm{init}}}\}_{\nu=0}^{\infty}$ is a complete set of the eigenspace of $N_r$ and $H_r$. We make the same observation for the other reservoirs and obtain the set of particles $\{Q_{Nr,\nu}^{\rm{init}}\}_{r=1}^{m}$ and energies  $\{Q_{Er,\nu}^{\rm{init}}\}_{r=1}^{m}$, which collapses the wave function of the whole system into  
\begin{equation}
\ket{\Psi_{\nu}^{\rm{init}}}=\bigotimes_{r=1}^{m}\ket{\psi_{r,\nu}^{\rm{init}}}\otimes \ket{\psi_{d}^{\rm{init}}},
\end{equation} 
where $\ket{\psi_{d}^{\rm{init}}}$ is the wave function of the quantum dots after measurement of all reservoirs. 
\item We let the whole system, which consists of $m$ reservoirs and $n$ quantum dots, evolve over time $s$ with the unitary operator $e^{-iHs}$.
\item At the time $t=s/2$, we again observe a reservoir $r$ to measure the number $Q_{Nr,\lambda}^{\rm{fin}}$ and the energy $Q_{Er,\lambda}^{\rm{fin}}$ of the particles in it, which collapses the wave function of each reservoir into the eigenfunction $\ket{\psi_{r,\lambda}^{\rm{fin}}}$ of the particle and energy operators:
\begin{align}
N_r\ket{\psi_{r,\lambda}^{\rm{fin}}} &= Q_{Nr,\lambda}^{\rm{fin}}\ket{\psi_{r,\lambda}^{\rm{fin}}}, \\
H_r\ket{\psi_{r,\lambda}^{\rm{fin}}} &= Q_{Er,\lambda}^{\rm{fin}}\ket{\psi_{r,\lambda}^{\rm{fin}}}.
\end{align}
Note that the set of kets $\{\ket{\psi_{r,\lambda}^{\rm{fin}}}\}_{\lambda=0}^{\infty}$ is also a complete set of the eigenspace of $N_r$ and $H_r$. We make the same observation for the other reservoirs and obtain the set of particles $\{Q_{Nr,\lambda}^{\mathrm{fin}}\}_{r=1}^{m}$ and energies $\{Q_{Er,\lambda}^{\rm{fin}}\}_{r=1}^{m}$, which collapses the wave function of the whole system into 
\begin{equation}
\ket{\Psi_\lambda^{\rm{fin}}}=\bigotimes_{r=1}^{m}\ket{\psi_{r,\lambda}^{\rm{fin}}}\otimes \ket{\psi_{d}^{\rm{fin}}}, 
\end{equation}
where $\ket{\psi_{d}^{\rm{fin}}}$ is the wave function of the quantum dots after measurement of all reservoirs. After the measurement, we obtain the conditional probability $P_{\nu\to \lambda}(\{Q_{Nr}\},\{Q_{Er}\})$ of finding the changes in the number $Q_{Nr}=Q_{Nr,\nu}-Q_{Nr,\lambda}$ and the energy $Q_{Er} =Q_{Er,\nu}-Q_{Er,\lambda}$ of the particles in each reservoir under the condition that the initial state of the whole system is fixed to $\ket{\Psi_\nu^{\rm{init}}}$:
\begin{align}
P_{\nu\to \lambda}(\{Q_{Nr}\},\{Q_{Er}\}) &=\left. \left|\braket{\Psi_\lambda^{\rm{fin}}|e^{-iHs}|\Psi_\nu^{\rm{init}}}\right|^2 \right|_{\nu,\lambda\ \text{s.t.}\  Q_{Nr}=Q_{Nr,\nu}-Q_{Nr,\lambda},Q_{Er} =Q_{Er,\nu}-Q_{Er,\lambda}} \notag\\
 &=  \sum_{\lambda=0}^{\infty} \left|\braket{\Psi_\lambda^{\rm{fin}}|e^{-iHs}|\Psi_\nu^{\rm{init}}}\right|^2 \notag \\
 & \quad \times \prod_{r=1}^{m}\delta[Q_{Nr} -(Q_{Nr,\nu}^{\rm{init}}-Q_{Nr,\lambda}^{\rm{fin}})]\delta[Q_{Er} -(Q_{Er,\nu}^{\rm{init}}-Q_{Er,\lambda}^{\rm{fin}})]. \label{eq:3.26}
\end{align}
\item We iterate the protocols 1 to 3 for the same initial density matrix $\rho_0$ and obtain the probability of finding the changes in the number $Q_{Nr}$ and the energy $Q_{Er}$ of the particles in each reservoir with the initial state $\ket{\Psi_\nu^{\rm{init}}}$ obeying the grand-canonical distribution $\braket{\Psi_\nu^{\rm{init}}|\rho_0|\Psi_\nu^{\rm{init}}}$ in the form: 
\begin{equation}
P(\{Q_{Nr}\},\{Q_{Er}\}) =\sum_{\nu,\lambda=0}^{\infty}P_{\nu\to \lambda}(\{Q_{Nr}\},\{Q_{Er}\})\braket{\Psi_\nu^{\rm{init}}|\rho_0|\Psi_\nu^{\rm{init}}}. \label{eq:3.27}
\end{equation}
\end{enumerate}
\subsection{Symmetry of the cumulant generating function}
We can calculate the characteristic function $Z(\chi)$ from the probability $P(\{Q_{Nr}\},\{Q_{Er}\})$ as follows: 
\begin{align}
Z(\{\chi_{cr}\},\{\chi_{hr}\};B) &= \sum_{\{Q_{Nr}\},\{Q_{Er}\}}P(\{Q_{Nr}\},\{Q_{Er}\})\prod_{r=1}^{m}e^{i(\chi_{cr}Q_{Nr}+\chi_{hr}Q_{Er})}  \label{eq:ichicQn} \\ 
&= \Tr{\rho_0\dagg{V}e^{iHs}V^2e^{-iHs}\dagg{V}}\notag \\
  & \equiv \braket{\dagg{V}e^{iHs}V^2e^{-iHs}\dagg{V}}_{\rho_0},\label{eq:chara}
\end{align}
where
\begin{equation}
V = \prod_{r=1}^{m}\exp\left[\frac{-i(\chi_{cr} N_r+\chi_{hr}H_r)}{2}\right]. \label{eq:Vdef}
\end{equation}
We show the details of the calculation in Appendix A.
This characteristic function has the following symmetry
\begin{equation}
Z(\{\chi_{cr}\},\{\chi_{hr}\};B) =Z(\{-\chi_{cr}+iA_{cr}\},\{-\chi_{hr}+iA_{hr}\};-B), \label{eq:Zsymm1}
\end{equation}
whose proof is also shown in Appendix A.

Using this characteristic function \eqref{eq:chara}, we can define the cumulant generating function of this model in the form:
\begin{equation}
F(\{\chi_{cr}\},\{\chi_{hr}\};B) = \lim_{s\to\infty}\frac{1}{s}\ln Z(\{\chi_{cr}\},\{\chi_{hr}\};B).
\end{equation}
The symmetry of the characteristic function \eqref{eq:Zsymm1} gives the symmetry of this cumulant generating function as follows:
\begin{equation}
F(\{\chi_{cr}\},\{\chi_{hr}\};B) =F(\{-\chi_{cr}+iA_{cr}\},\{-\chi_{hr}+iA_{hr}\};-B). \label{eq:Fsymm1}
\end{equation}
Noting that the cumulant generating function only depends on the difference between two counting fields \cite{saito2008} as in 
\begin{equation}
F(\{\chi_{cr}\},\{\chi_{hr}\};B) = F(\{\bm{\chi_{cr}}\},\{\bm{\chi_{hr}}\};B),
\end{equation}
where $\bm{\chi_{cr}} = \chi_{cr}-\chi_{cr'}$ and $\bm{\chi_{hr}} = \chi_{hr}-\chi_{hr'}$ with $r'$ fixed and $r' \neq r$,
we obtain the following symmetry of the cumulant generating function from the symmetry \eqref{eq:Fsymm1}:
\begin{equation}
F(\{\bm{\chi_{cr}}\},\{\bm{\chi_{hr}}\};B) =F(\{-\bm{\chi_{cr}}+i\bm{A_{cr}}\},\{-\bm{\chi_{hr}}+i\bm{A_{hr}}\};-B),
\label{eq:Fsymm} 
\end{equation}
where $\bm{A_{cr}}$ and $\bm{A_{hr}}$ are the affinities (thermodynamic forces) of the particle flow and the energy flow, respectively:
\begin{align}
\bm{A_{cr}} = \beta_r\mu_r-\beta_{r'}\mu_{r'}, \label{eq:Acr}\\
\bm{A_{hr}} = -\beta_r+\beta_{r'}, \label{eq:Ahr}
\end{align}
with $r'$ fixed to one of the affinities and $r \neq r'$.
We will explain why we choose these affinities in Chap.~4.
The symmetry \eqref{eq:Fsymm} produces many interesting relations of transport coefficients. We next see the relations for the simplest two-terminal case.
\subsection{Generalized Onsager relation in the case of two terminals}   
Let us consider the case of two terminals, in which we can write the symmetry of the cumulant generating function in the form:
\begin{equation}
F(\chi_{c},\chi_{h};B) =F(-\chi_{c}+iA_{c},-\chi_{h}+iA_{h};-B), \label{eq:Fsymm2} 
\end{equation} 
where $\chi_c =\chicL-\chicR$, $\chi_h =\chihL-\chihR$, $A_c =\AcL-\AcR = \betaL\muL-\betaR\muR$, and $A_h =\AhL-\AhR=-\betaL+\betaR$.
We can compute the $k$th cumulant of the particle flow $J_N$ and the energy flow $J_E$ by differentiating the cumulant generating function with respect to $\chi_c$ and $\chi_h$ as follows:
\begin{equation}
\cumulant{J_N^{k_1}J_E^{k_2}} = \left. \frac{\partial^{k_1+k_2}F(\chi_c,\chi_h;B)}{\partial (i\chi_{c})^{k_1}\partial (i\chi_{h})^{k_2}} \right|_{\chi_c=\chi_h=A_c=A_h=0}.
\end{equation}
We then define the transport coefficients as follows:
\begin{equation}
L^{k_1,k_2}_{\ell_1,\ell_2}(B) = \frac{\partial^{\ell_1+\ell_2}\cumulant{J_N^{k_1}J_E^{k_2}}}{\partial A_{c}^{\ell_1}\partial A_{h}^{\ell_2}}.
\end{equation}

For later use, we symmetrize and antisymmetrize the transport coefficients and the cumulant generating function with respect to the magnetic field:
\begin{align}
F_{\pm}(\chi_{c},\chi_{h};B) &= F(\chi_{c},\chi_{h};B)\pm F(\chi_{c},\chi_{h};-B), \\
L^{k_1,k_2}_{\ell 1,\ell 2 \pm}(B) &= L^{k_1,k_2}_{\ell_1,\ell_2}(B)\pm L^{k_1,k_2}_{\ell_1,\ell_2}(-B),
\end{align}
which satisfy the relation
\begin{equation}
L^{k_1,k_2}_{\ell_1,\ell_2\pm}(B) = \left. \frac{\partial^{k_1+k_2+\ell_1+\ell_2}F_{\pm}(\chi_c,\chi_h;B)}{\partial(i\chi_{c})^{k_1}\partial(i\chi_{h})^{k_2}\partial A_{c}^{\ell_1}\partial A_{h}^{\ell_2}}\right|_{\chi_c=\chi_h=A_c=A_h=0}.
\end{equation}
The symmetry in Eq.~\eqref{eq:Fsymm2} gives the symmetry of $F_{\pm}(\chi_{c},\chi_{h};B)$ in the form:
\begin{equation}
F_{\pm}(\chi_{c},\chi_{h};B) = \pm F_{\pm}(-\chi_{c}+iA_{c},-\chi_{h}+iA_{h};B). \label{eq:Fpmsymm}
\end{equation}

Partially differentiating both sides of Eq.~\eqref{eq:Fpmsymm} with respect to $\chi_{c}$ $k_1$ times, $\chi_{h}$ $k_2$ times, $A_{c}$ $\ell_1$ times, and $A_{h}$ $\ell_2$ times, we arrive at the relation of the transport coefficients as follows:
\begin{align}
&\left. \frac{\partial^{k_1+k_2+\ell_1+\ell_2}F_{\pm}(\chi_{c},\chi_{h};B)}{\partial(i\chi_{c})^{k_1}\partial(i\chi_{h})^{k_2}\partial A_{c}^{\ell_1}\partial A_{h}^{\ell_2}}\right|_{\chi_{c}=\chi_{h}=A_{c}=A_{h}=0}  \notag \\
 &= \pm\left. \frac{\partial^{k_1+k_2+\ell_1+\ell_2}F_{\pm}(-\chi_{c}+iA_{c},-\chi_{h}+iA_{h};B)}{\partial(i\chi_{c})^{k_1}\partial(i\chi_{h})^{k_2}\partial A_{c}^{\ell_1}\partial A_{h}^{\ell_2}}\right|_{\chi_{c}=\chi_{h}=A_{c}=A_{h}=0},
\end{align}
which produces
\begin{equation}
L^{k_1,k_2}_{\ell_1,\ell_2\pm}(B) = \pm \sum_{n_1=0}^{\ell_1}\sum_{n_2=0}^{\ell_2}\binom{\ell_1}{n_1}\binom{\ell_2}{n_2}(-1)^{n_1+n_2+k_1+k_2}L^{k_1+n_1, k_2+n_2}_{\ell_1-n_1,\ell_2-n_2\pm}(B). \label{eq:GeneL}
\end{equation} 
Note that the cumulant generating function $F(\chi_{c},\chi_{h};B)$ also depends on the affinities $A_c$ and $A_h$ when one differentiates $F(\chi_{c},\chi_{h};B)$. The relation \eqref{eq:GeneL} is among the coefficients with a fixed value of $N=k_1+k_2+\ell_1+\ell_2$. For $N=2$, we have 
\begin{align}
L^{01}_{10}(B) &= L^{10}_{01}(-B), \label{eq:L0110}\\
L^{10}_{10}(B) &= L^{10}_{10}(-B), \label{eq:L1010}\\
L^{01}_{01}(B) &= L^{01}_{01}(-B),
\end{align}
which are known as the Onsager-Casimir relations, as well as
\begin{align}
L^{20}_{00}(B) &= 2L^{10}_{10}(B), \label{eq:L2000} \\
L^{02}_{00}(B) &= 2L^{01}_{01}(B),
\end{align}
which are known as the fluctuation-dissipation theorem. 
For $N=3$, we have
\begin{align}
L^{20}_{10,+}(B) &= L^{10}_{20,+}(B), \label{eq:L2010+} \\
L^{11}_{10,+}(B) &= L^{01}_{20,+}(B) = 2L^{10}_{11,+}(B)-L^{20}_{01,+}(B), \\
L^{11}_{01,+}(B) &= L^{10}_{02,+}(B) = 2L^{01}_{11,+}(B)-L^{02}_{10,+}(B), \\
L^{02}_{01,+}(B) &= L^{01}_{02,+}(B), \\
L^{30}_{00,+}(B) &= L^{03}_{00,+}(B), \\
L^{10}_{20,-}(B) &= \frac{L^{20}_{10,-}(B)}{3}=\frac{L^{30}_{00,-}(B)}{6}, \label{eq:L1020-}\\
L^{01}_{02,-}(B) &=  \frac{L^{02}_{01,-}(B)}{3}=\frac{L^{03}_{00,-}(B)}{6}, \\
L^{20}_{01,-}(B) &= L^{01}_{20,-}(B)+2L^{10}_{11,-}(B), \\
L^{02}_{10,-}(B) &= L^{10}_{02,-}(B)+2L^{01}_{11,-}(B).\label{eq:L0210-}
\end{align}
The relations among the transport coefficients of the cumulant of the particle flow \eqref{eq:L1010}, \eqref{eq:L2000}, \eqref{eq:L2010+}, and \eqref{eq:L1020-}  have been observed experimentally \cite{nakamura2010, nakamura2011}.

\chapter{The Generalized Onsager Relations between Heat Flow and Particle Flow}
In the previous chapter, we consider the higher-order cumulants of the particle and energy flows. 
In this chapter, we treat the higher-order cumulants of the \textit{heat} flow.
We show the appropriate definition of the heat flow in mesoscopic transport systems and derive the generalized Onsager relations for the particle and heat flows.  
In Section 4.1, we review the several definitions of the heat flows, which might have confused researches in the past. 
We then explain the heat flow which seems to be the most relevant one. 
In Section 4.2, we first explain the affinities. 
We then derive the generalized Onsager relations for the particle and heat flows when we expand these flows with respect to the appropriate affinities. We also consider the generalized flows which are superpositions of the particle and energy flows. 
We show that we can choose the appropriate affinities of the generalized flows and derive the relations among transport coefficients of the flows when we expand them with respect to their affinities. 
For simplicity, we treat only the setup with two reservoirs throughout this chapter.
\section{Heat flow in mesoscopic transport systems}
\subsection{Several `heat' flows in mesoscopic transport systems}
We here introduce three expressions of the `heat' flow in mesoscopic transport systems. The thermodynamic definition of heat is clear, but the wording of  the `heat flow' may be confusing because there seems to be several definitions. 

The first definition of the heat flow is given by 
\begin{equation}
J_E = \frac{1}{h} \int_{\EG}^{\infty}d\epsilon \tau(\epsilon) \epsilon(\fL(\epsilon)-\fR(\epsilon)), \label{eq:1def}
\end{equation}
which is used widely in researches of mesoscopic transport~\cite{iyoda2010,ruokola2011,lim2013}. Although this `heat' flow should be called an `energy' flow, it is called a `heat' flow probably because it is considered in the situation where the energy flow does not do work and hence all energy becomes heat.

The second definition of the heat flow is given by
\begin{equation}
J_Q = \frac{1}{h} \int_{\EG}^{\infty}d\epsilon \tau(\epsilon) (\epsilon-\mu)(\fL(\epsilon)-\fR(\epsilon)), \label{eq:2def}
\end{equation}  
which was often used in the dawn of the research of heat flow in mesoscopic systems~\cite{sivan}. 
This definition may have been taken from an equation in Callen's textbook~\cite{callen},
\begin{equation}
J_Q = J_E - \mu J_N, \label{eq:callen}
\end{equation}   
where $J_E$ is the energy flow given by Eq.~(\ref{eq:1def}), while $J_N$ is the particle flow, for which the Landauer-B\"uttiker formalism gives the expression 
\begin{equation}
J_N = \frac{1}{h} \int_{\EG}^{\infty}d\epsilon \tau(\epsilon)(\fL(\epsilon)-\fR(\epsilon)). \label{eq:jn}
\end{equation}
Substituting Eq.~(\ref{eq:1def}) and Eq.~(\ref{eq:jn}) into Eq.~(\ref{eq:callen}), we would obtain the second definition (\ref{eq:2def}).

Since this `heat' flow was not microscopically derived, we do not know clearly where it flows. It is indeed ambiguous of which part of the system in Fig.~\ref{fig:system0} the chemical potential $\mu$ of Eq.~\eqref{eq:callen} is. We should probably choose $\mu$ so that $J_Q$ may satisfy Onsager's reciprocal theorem. For example, in Ref.~\cite{sivan}, the authors chose $\mu$ as $(\muL+\muR)/2$ and in Ref.~\cite{Butcher}, the author chose $\mu$ as $\muL$. The choices do not make difference in the linear response of the voltage difference $(\muR-\muL)/e$ but differ in higher orders.

The third definition is given by
\begin{equation}
J_Q^{\alpha} = \frac{1}{h} \int_{\EG}^{\infty}d\epsilon \tau(\epsilon) (\epsilon-\mu_{\alpha})(\fL(\epsilon)-\fR(\epsilon)), \label{eq:3def}
\end{equation}
where $\alpha=\mathrm{L},\mathrm{R}$.
This appears to be the same as the definition (\ref{eq:jn}) with the chemical potential arbitrarily fixed, but we here make distinction because the definition (\ref{eq:3def})
clearly specifies where the heat flow exists; we will show below that this definition gives the heat flow out of the left reservoir or into the right reservoir. 
\subsection{Definition of the heat flow}
In this section, we derive the third definition of the heat flow \eqref{eq:3def}
using the Landauer-B\"{u}ttiker formula and thermodynamics. We here derive $\JQL$ specifically, but we can derive $\JQR$ in the same way. 

\begin{figure}[h]
\begin{center}
\includegraphics[width=10cm]{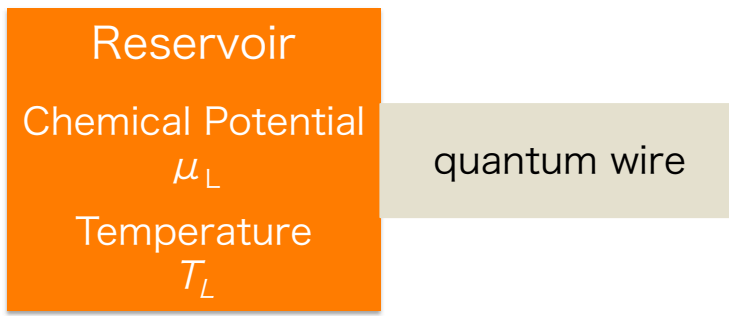}
\end{center}
\caption{The wire and the left reservoir.}
\label{fig:derive}
\end{figure}

Consider the quantum wire with the left reservoir (Fig.~\ref{fig:derive}). We assume that the reservoir is so large that it is always in equilibrium. Hence we can define thermodynamic quantities of the reservoir such as the temperature and the chemical potential. 

Since each reservoir is coupled only with the wire, all the heat generated in the  left reservoir goes into the wire. Therefore, 
\begin{equation}
\dot{\QL} = -\JQL, \label{eq:QL1}
\end{equation}
where $\dot{\QL}$ is the heat generated in the left reservoir per unit time. The negative sign appears because we define the positive direction so that flows going to the right may be positive. 

Using the first law of thermodynamics, $\mathrm{d}Q=\mathrm{d}U+\mathrm{d}W$, we obtain 
\begin{equation}
\dot{\QL} = \dot{\UL} - \dot{\WL}, \label{eq:QL2}
\end{equation}
where $\dot{\UL}$ is the energy going in the left reservoir per unit time and $\dot{\WL}$ is the work per unit time done on the left reservoir. Let us use the framework in which the volume of the reservoir is constant. Hence, the pressure dose not do work on the reservoir. Therefore,
\begin{equation}
\dot{\WL} = \muL \dot{\NL}, \label{eq:WL}
\end{equation}
where $\dot{\NL}$ is the number of particles going in the left reservoir per unit time. 

In order to find $\dot{\UL}$ and $\dot{\NL}$ microscopically, we use the Landauer-B\"uttiker formalism, which gives
\begin{align}
 \dot{\UL} &= -\frac{1}{h}\int_{\EG}^{\infty} d\epsilon \tau(\epsilon)\epsilon(\fL(\epsilon)-\fR(\epsilon)), \label{eq:UL} \\
 \dot{\NL} &= -\frac{1}{h}\int_{\EG}^{\infty} d\epsilon  \tau(\epsilon)(\fL(\epsilon)-\fR(\epsilon)). \label{eq:NL}
\end{align}
We thus arrive at 
\begin{equation}
\JQL = \frac{1}{h}\int_{\EG}^{\infty} d\epsilon  \tau(\epsilon)(\epsilon-\muL)(\fL(\epsilon)-\fR(\epsilon)). \label{eq:jql}
\end{equation}
We can derive $\JQR$ in the same way, expect that  we have $\dot{\QR} = \JQR$ and therefore the sign of $\dot{\UR}$ and $\dot{\NR}$ are reversed, which results in
\begin{equation}
\JQR = \frac{1}{h}\int_{\EG}^{\infty} d\epsilon  \tau(\epsilon)(\epsilon-\muR)(\fL(\epsilon)-\fR(\epsilon)).\label{eq:jqr}
\end{equation}

These heat flows $\JQL$ and $\JQR$ have not been observed experimentally as far as we know. In order to  observe them, the experimental condition should be at low temperatures as was when the energy flow  was observed \cite{jezouin2013}. We therefore show in appendix B the expansion of $\JQL$ and $\JQR$ at low temperatures. 
\subsection{Constructing heat engine and its efficiency with the heat flow}

\begin{figure}[h]
\begin{center}
\includegraphics[width=13cm]{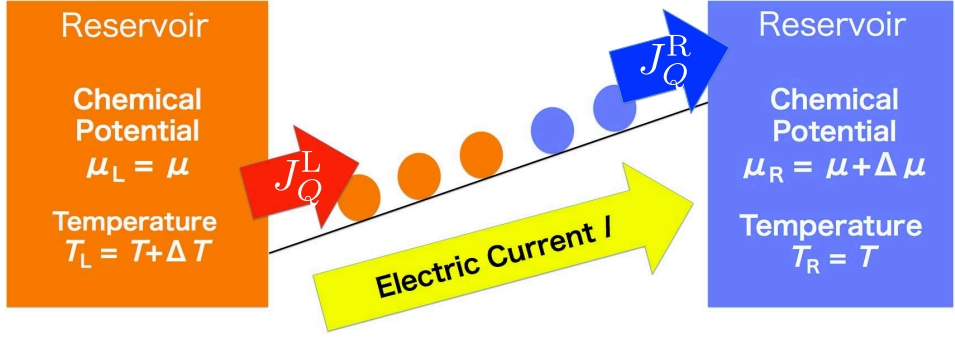}
\end{center}
\caption{Schematic picture of a mesoscopic heat engine.
Note that we cannot define the chemical potential in the wire as this illustration might suggest because it is highly non-equilibrium there and therefore the thermodynamic quantities, such as the chemical potential cannot be defined.}
\label{fig:engine}
\end{figure}

The heat flows $\JQL$ and $\JQR$ in Eq.~\eqref{eq:3def} can be used to analyze mesoscopic heat engines~\cite{Benenti} in Fig.~\ref{fig:engine}.
We set the chemical potential of the right reservoir higher than the left, while the temperature of the left reservoir higher than the right so that an electric current may go from left to right against the difference of the chemical potential. 
What happens per unit time is the following. Electrons gain heat $\JQL$ from the hot left reservoir, go to the right against the potential difference, during which electrons do the work of amount $IV$, where $I$ is the electric current and $V$ is the voltage difference, and then dump heat $\JQR$ to the cold right reservoir. We can thus consider this system as a heat engine. 
Its efficiency  $\eta$ is thereby given by
\begin{equation}
\eta = \frac{IV}{\JQL}. \label{eq:effc}
\end{equation}
 
The current $I$ is given by the Landauer-B\"uttiker formula
\begin{equation}
I= \frac{e}{h}\int_{\EG}^{\infty} d\epsilon \tau(\epsilon)(\fL(\epsilon)-\fR(\epsilon)),
\end{equation}
while the voltage difference is given by
\begin{equation}
V = \frac{\muR-\muL}{e}.
\end{equation}
Therefore, the work $IV$ in Eq.~(\ref{eq:effc}) is given by 
\begin{align}
 IV &= \frac{\muR-\muL}{e}\cdot \frac{e}{h}\int_{\EG}^{\infty} d\epsilon \tau(\epsilon)(\fL(\epsilon)-\fR(\epsilon)) \\
 &= \frac{1}{h}\int_{\EG}^{\infty} d\epsilon  \tau(\epsilon)(\epsilon-\muL) (\fL(\epsilon)-\fR(\epsilon))-\frac{1}{h}\int_{\EG}^{\infty} d\epsilon \tau(\epsilon) (\epsilon-\muR)(\fL(\epsilon)-\fR(\epsilon)) \\
     &= \JQL - \JQR,
\end{align}
which results in 
\begin{equation}
\eta = 1- \frac{\JQR}{\JQL}.
\end{equation}
We show in Appendix C that the upper limit of this efficiency is the Carnot efficiency as is expected from the theory of the standard heat engine.
\subsection{Affinity}
\subsubsection{Definition}
Affinities, or thermodynamic forces are the forces which drive a system in equilibrium out of it, such as the difference in the temperature, the chemical potential, the pressure, and so on.
We define the affinities as follows. We assume that the entropy of the whole system is a function of a set of extensive variables $\{X_k\}$:
\begin{equation}
S = S(X_0,X_1,X_2, \cdots).
\end{equation}
The entropy production of the whole system is thus given by
\begin{equation}
\Sdot \equiv \frac{\mathrm{d}S}{\mathrm{d}t} = \sum_{k}\frac{\partial S}{\partial X_k}\frac{\mathrm{d}X_k}{\mathrm{d}t} \equiv \sum_{k} J_kA_k, 
\end{equation} 
where we define the affinity $A_k$ and the corresponding flux $J_k$ by
\begin{align}
A_k &\equiv \frac{\partial S}{\partial X_k} ,\\
J_k &\equiv \frac{\mathrm{d}X_k}{\mathrm{d}t}. 
\end{align}

Let us generalize this definition of affinities and fluxes as follows. When the entropy production can be expressed as a sum of the products of an extensive flux and an intensive parameter,
\begin{equation}
\Sdot = \sum_{k} J_kA_k,
\end{equation}
we call $A_k$ an affinity and $J_k$ the corresponding flux. We show below  examples of $A_k$ and $J_k$.
\subsubsection{Example 1: Affinities corresponding to $\JQL$ and $\JQR$}
Let us consider the setup with two terminals shown in Fig.~\ref{fig:system}. As no entropy is generated in the wire, the total entropy production is given by the sum of the entropy production of each reservoir: 
\begin{align}
\Sdot &= \SLdot + \SRdot \notag \\
 &= \betaL\QLdot + \betaR \QRdot \notag \\
 &= -\betaL\JQL + \betaR \JQR \notag \\
 &\equiv \AqL\JQL + \AqR\JQR,
\end{align}
which lets us define the affinities $\AqL \equiv -\betaL$ and $\AqR \equiv \betaR$ of the flows $\JQL$ and $\JQR$, respectively. Note that the affinities are not a difference of intensive parameters in this case, which might be because $\JQL$ and $\JQR$ are not conserved quantities. 
\subsubsection{Example 2: Affinities corresponding to $J_E$ and $J_N$}
We here explain why we chose the affinities of $J_E$ and $J_N$ as Eqs.~\eqref{eq:Acr} and \eqref{eq:Ahr}. Using the relations
$\JQL = J_E - \muL J_N$ and $\JQR = J_E - \muR J_N$, 
we can transform the entropy production as follows:
\begin{align}
\Sdot &=-\betaL\JQL + \betaR \JQR \notag \\
 &= -\betaL(J_E - \muL J_N) + \betaR(J_E - \muR J_N) \notag \\
 &= (-\betaL+\betaR)J_E + (\betaL\muL-\betaR\muR)J_N \notag \\
 & \equiv \Ah J_E + \Ac J_N,  
\end{align}
which leads to the definitions of the affinities $\Ah \equiv -\betaL+\betaR$ and $\Ac \equiv \betaL\muL-\betaR\muR$ of the flows $J_E$ and $J_N$, respectively.
\subsubsection{Example 3: Affinities corresponding to $\JQL$ or $\JQR$ and $J_N$}
We then derive the affinities corresponding to $\JQL$ and $J_N$. Using the relation
$\JQL-\JQR = (\muR-\muL)J_N$,
we can transform the entropy production as follows:
\begin{align}
\Sdot &=  -\betaL\JQL + \betaR \JQR \notag \\
 &= -\betaL\JQL + \betaR[\JQL-(\muR-\muL)J_N] \notag \\
 &= (-\betaL+\betaR)\JQL + \betaR(\muL-\muR)J_N \notag \\
 &\equiv \AQL\JQL + \ANL J_N,
\end{align}
which leads to $\AQL \equiv -\betaL+\betaR$ and $\ANL \equiv \betaR(\muL-\muR)$ of the flows $\JQL$ and $J_N$, respectively. 
In the same manner, we can find the affinities corresponding to $\JQR$ and $J_N$:
\begin{equation}
\Sdot = \AQR\JQR + \ANR J_N,
\end{equation}
where 
$\AQR \equiv -\betaL+\betaR$ and $\ANR \equiv \betaL(\muL-\muR)$.

The notable point here is that the second affinity in either case is proportional to the difference in the chemical potential. This was not the case in Example 2.
\subsubsection{Example 4: Affinities corresponding to flows which are superpositions of $J_E$ and $J_N$}
We here consider flows which are general superpositions of $J_E$ and $J_N$. Let us define the generalized flows $J_1$ and $J_2$ by
\begin{equation}
\begin{pmatrix}
J_1 \\
J_2
\end{pmatrix}
=
K
\begin{pmatrix}
J_E \\
J_N
\end{pmatrix}, \label{eq:J1J2}
\end{equation}
where
\begin{equation}
K \equiv \begin{pmatrix}
a&b \\
c&d
\end{pmatrix},
\end{equation}
with the condition that $a$, $b$, $c$, $d$ are real and $\mathrm{det}K = ad-bc \neq 0$.
The affinities of these flows are given by
\begin{equation}
\begin{pmatrix}
A_1 &A_2
\end{pmatrix}
=
\begin{pmatrix}
\Ah & \Ac
\end{pmatrix}
K^{-1},
\end{equation}
because we then have
\begin{align}
\Sdot &= \begin{pmatrix} \Ah &\Ac \end{pmatrix} \begin{pmatrix} J_E \\ J_N \end{pmatrix} \notag \\
   &= \begin{pmatrix} \Ah & \Ac \end{pmatrix}K^{-1} K \begin{pmatrix} J_E \\ J_N \end{pmatrix} \notag \\
   &= \begin{pmatrix} A_1 &A_2 \end{pmatrix} \begin{pmatrix} J_1 \\ J_2 \end{pmatrix} \notag \\
   &= A_1J_1 + A_2J_2. \label{eq:SdotJ1J2}
\end{align}

For Example 1, the matrix $K$ is given by
\begin{equation}
K= \begin{pmatrix} 1&-\muL \\ 1&-\muR \end{pmatrix}
\end{equation}
as in
\begin{equation}
\begin{pmatrix}
\JQL \\
\JQR
\end{pmatrix}
=
K
\begin{pmatrix}
J_E \\
J_N
\end{pmatrix}, \label{eq:JQLJQR}
\end{equation}
while for Example 3, it is
\begin{equation}
K= \begin{pmatrix} 1&-\muL \\ 0&1 \end{pmatrix}
\end{equation}
as in
\begin{equation}
\begin{pmatrix}
\JQL \\
J_N
\end{pmatrix}
=
K
\begin{pmatrix}
J_E \\
J_N
\end{pmatrix}. \label{eq:JQLJN}
\end{equation}
We thereby confirm
\begin{align}
\begin{pmatrix} \AqL & \AqR \end{pmatrix} &= \begin{pmatrix} A_h & A_c  \end{pmatrix} \begin{pmatrix} 1&-\muL \\ 1&-\muR \end{pmatrix}^{-1} \notag \\
&=\frac{1}{\muL-\muR} \begin{pmatrix} -\betaL+\betaR &\betaL\muL-\betaR\muR  \end{pmatrix} \begin{pmatrix} -\muR&\muL \\ -1&1 \end{pmatrix} \notag \\
&= \begin{pmatrix} -\betaL & \betaR  \end{pmatrix},
\end{align}
and
\begin{align}
\begin{pmatrix} \AQL & \ANL \end{pmatrix} &= \begin{pmatrix} A_h & A_c  \end{pmatrix} \begin{pmatrix} 1&-\muL \\ 0&1 \end{pmatrix}^{-1} \notag \\
&=\begin{pmatrix} -\betaL+\betaR &\betaL\muL-\betaR\muR  \end{pmatrix} \begin{pmatrix} 1&\muL \\ 0&1 \end{pmatrix} \notag \\
&= \begin{pmatrix} -\betaL+\betaR & \betaR(\muL-\muR)  \end{pmatrix}.
\end{align}
\subsection{Counting fields}
Let us consider the counting fields corresponding to the generalized flows $J_1$ and $J_2$. We transform the counting fields $\chi_c$ and $\chi_h$ so that the exponent in Eq.~\eqref{eq:ichicQn},  
\begin{equation}
\chi_cQ_N+\chi_hQ_E = \begin{pmatrix} \chi_h & \chi_c \end{pmatrix} \begin{pmatrix} Q_E \\ Q_N \end{pmatrix},
\end{equation}
may not change. We can achieve it by defining the new counting fields $\chi_1$, $\chi_2$ by
\begin{equation}
\begin{pmatrix}
\chi_1 & \chi_2
\end{pmatrix}
=
\begin{pmatrix}
\chi_h & \chi_c
\end{pmatrix}K^{-1}.
\end{equation}
Noting that $Q_1$ and $Q_2$ are transformed as $J_1$ and $J_2$, we can show that the transformation does not change exponent in Eq.~\eqref{eq:ichicQn}:
\begin{align}
\chi_cQ_N+\chi_hQ_E &= \begin{pmatrix} \chi_h & \chi_c \end{pmatrix} \begin{pmatrix} Q_E \\ Q_N \end{pmatrix} \notag \\
 &=  \begin{pmatrix} \chi_h & \chi_c \end{pmatrix}K^{-1} K \begin{pmatrix} Q_E \\ Q_N \end{pmatrix} \notag \\
 &=  \begin{pmatrix} \chi_1 & \chi_2 \end{pmatrix} \begin{pmatrix} Q_1 \\ Q_2 \end{pmatrix} \notag \\
 &= \chi_1Q_1+\chi_2Q_2.
\end{align}

Let us present the counting fields of several flows;
for $\JQL$ and $\JQR$ we have \eqref{eq:JQLJQR},
and their counting fields $\chiqL$ and $\chiqR$ are therefore given by
\begin{equation}
\begin{pmatrix}
\chiqL & \chiqR
\end{pmatrix}
=
\begin{pmatrix}
\chi_h & \chi_c
\end{pmatrix}
\begin{pmatrix}
1&-\muL \\
1&-\muR
\end{pmatrix}^{-1};
\end{equation}
for $\JQL$ and $J_N$, because of \eqref{eq:JQLJN},
the counting fields $\chiQL$ and $\chiNL$ are given by
\begin{equation}
\begin{pmatrix}
\chiQL & \chiNL
\end{pmatrix}
=
\begin{pmatrix}
\chi_h & \chi_c
\end{pmatrix}
\begin{pmatrix}
1&-\muL \\
0&1
\end{pmatrix}^{-1}.
\end{equation}
We can define the counting fields for $\JQR$ and $J_N$ in the same manner.
\section{The generalized Onsager relations between the heat and particle flows}
In this section, we derive the generalized Onsager relations between several pairs of flows.
\subsection{General discussion}
Let us define the cumulant generating function for $J_1$ and $J_2$ by
\begin{equation}
F_{J_1J_2}(\chi_1,\chi_2,B) \equiv F(\chi_c,\chi_h,B)=F(b\chi_{1}+d\chi_2,a\chi_{1}+c\chi_2;B), 
\end{equation}
with which we can define the transport coefficients of $J_1$ and $J_2$ as follows:
\begin{equation}
\mathcal{L}^{k_1,k_2}_{\ell_1,\ell_2}(B) = \left.\frac{\partial^{k_1+k_2+\ell_1+\ell_2}F_{J_1J_2}(\chi_1,\chi_2;B)}{\partial(i\chi_{1})^{k_1}\partial(i\chi_{2})^{k_2}\partial A_{1}^{\ell_1}\partial A_{2}^{\ell_2}}\right|_{\chi_1=\chi_2=A_1=A_2=0}. \label{eq:mathL}
\end{equation}
We can prove the following symmetry:
\begin{equation}
F_{J_1J_2\pm}(\chi_{1},\chi_{2};B) = \pm F_{J_1J_2\pm}(-\chi_{1}+iA_{1},-\chi_{2}+iA_{2};B), \label{eq:Genesymm}
\end{equation}
where 
\begin{equation}
F_{J_1J_2\pm}(\chi_{1},\chi_{2};B) = F_{J_1J_2}(\chi_{1},\chi_{2};B)\pm F_{J_1J_2}(\chi_{1},\chi_{2};-B). \label{eq:F12symm}
\end{equation}
The proof is as follows:
\begin{align}
F_{J_1J_2\pm}(\chi_{1},\chi_{2};B) &\equiv F_{\pm}(b\chi_{1}+d\chi_2,a\chi_{1}+c\chi_2;B) \notag \\
 &= F_{\pm}(\chi_{c},\chi_{h};B) \notag \\
 &= \pm F_{\pm}(-\chi_{c}+iA_{c},-\chi_{h}+iA_{h};B) \notag \\
 &=  \pm F_{\pm}[b(-\chi_{1}+iA_{1})+d(-\chi_2+iA_{2}),a(-\chi_{1}+iA_{1})+c(-\chi_2+iA_{2});B] \notag \\
 &=  \pm F_{J_1J_2\pm}(-\chi_{1}+iA_{1},-\chi_{2}+iA_{2};B),
\end{align}
where we used the symmetry \eqref{eq:Fsymm2} in the third equality.

From this symmetry \eqref{eq:F12symm}, we derive the relations among the transport coefficients using the same procedure as in Eqs.~\eqref{eq:Fpmsymm}--\eqref{eq:GeneL}:
\begin{equation}
\mathcal{L}^{k_1,k_2}_{\ell_1,\ell_2\pm}(B) = \pm \sum_{n_1=0}^{\ell_1}\sum_{n_2=0}^{\ell_2}\binom{\ell_1}{n_1}\binom{\ell_2}{n_2}(-1)^{n_1+n_2+k_1+k_2}\mathcal{L}^{k_1+n_1, k_2+n_2}_{\ell_1-n_1,\ell_2-n_2\pm}(B), \label{eq:mathLrela}
\end{equation}
where 
\begin{equation}
\mathcal{L}^{k_1,k_2}_{\ell_1,\ell_2\pm}(B) =\mathcal{L}^{k_1,k_2}_{\ell_1,\ell_2}(B) \pm\mathcal{L}^{k_1,k_2}_{\ell_1,\ell_2}(-B). 
\end{equation}
There are an infinite number of choices of $J_1$ and $J_2$, and hence we can derive an infinite number of corresponding Onsager relations. Of course, not all the choices are physically relevant. We show below examples of $J_1$ and $J_2$ which have physical meaning. 
\subsection{The generalized Onsager relations between $\JQL$ and $\JQR$}
Let us first consider the case $J_1=\JQL$ and $J_2=\JQR$. Because we have
\begin{align}
\begin{pmatrix} \chi_h & \chi_c \end{pmatrix} &= \begin{pmatrix} \chiqL & \chiqR \end{pmatrix}K \notag \\
&=
\begin{pmatrix} \chiqL & \chiqR \end{pmatrix}
\begin{pmatrix}
1&-\muL \\
1&-\muR
\end{pmatrix} \notag \\
&=
\begin{pmatrix}
\chiqL + \chiqR & -\muL\chiqL-\muR\chiqR 
\end{pmatrix},
\end{align}
the cumulant generating function for $\JQL$ and $\JQR$ is given by 
\begin{equation}
F_{\JQL\JQR}(\chiqL,\chiqR,B) \equiv F(-\muL\chiqL-\muR\chiqR,\chiqL+\chiqR;B),
\end{equation}
with which we can define the transport coefficients of $\JQL$ and $\JQR$  in the form \eqref{eq:mathL}. The symmetry \eqref{eq:Genesymm} and the relations \eqref{eq:mathLrela} follow from it.
\subsection{The generalized Onsager relations between $\JQL$ or $\JQR$ and $J_N$}
We next consider the case $J_1 = \JQL$ and $J_2=J_N$. Because we have
\begin{align}
\begin{pmatrix} \chi_h & \chi_c \end{pmatrix} &= \begin{pmatrix} \chiQL & \chiNL \end{pmatrix}
\begin{pmatrix}
1&-\muL \\
0&1
\end{pmatrix} \notag \\
&=
\begin{pmatrix}
\chiQL & -\muL\chiQL+\chiNL 
\end{pmatrix},
\end{align}
the cumulant generating function for $\JQL$ and $J_N$ is given by 
\begin{equation}
F_{\JQL J_N}(\chiQL,\chiNL,B) \equiv F(-\muL\chiQL+\chiNL,\chiQL;B), \label{eq:FJQLJNsymm}
\end{equation}
with which we can define the transport coefficients of $\JQL$ and $J_N$ in the form \eqref{eq:mathL}. We will specifically use the notation $G$ for the present case instead of $\mathcal{L}$ hereafter for the use in Subsection 4.2.4.
The symmetry \eqref{eq:Genesymm} and the relations \eqref{eq:mathLrela} again follow from \eqref{eq:FJQLJNsymm}. We remark that the relations of $G$ reproduce Eqs.~\eqref{eq:GNN}--\eqref{eq:GQQ} by identifying $G^{10}_{10} = G_{NN}$, $G^{10}_{01} = G_{NQ}$, $G^{01}_{10} = G_{QN}$, and $G^{01}_{01} = G_{QQ}$. However, we need to consider the present treatment in order to find the relations.

We can do the same for the case $J_1=\JQR$ and $J_2=J_N$, for which we will use the notation $M$ instead of $\mathcal{L}$. We note here that the linear coefficients of $M$ coincide with those of $G$, but they differ in higher orders.
\subsection{Application: Nonlinear Seebeck coefficient}
We here show that we can express the nonlinear Seebeck coefficient easily using the transport coefficients $G^{k_1,k_2}_{\ell_1,\ell_2}(B)$ or $M^{k_1,k_2}_{\ell_1,\ell_2}(B)$. We will use $G^{k_1,k_2}_{\ell_1,\ell_2}(B)$ hereafter, but we can use $M^{k_1,k_2}_{\ell_1,\ell_2}(B)$ in the same way. The advantage to use the coefficients $G^{k_1,k_2}_{\ell_1,\ell_2}(B)$ or $M^{k_1,k_2}_{\ell_1,\ell_2}(B)$ is that the corresponding affinities $\ANL$ or $\ANR$ and $A_N$ contain $\Delta \mu$ and $\Delta T$ explicitly in contrast to the affinities $A_c=\betaL\muL-\betaR\muR$. This enables us to expand thermoelectric coefficients, which are usually related to $\Delta \mu$ or $\Delta T$, more easily.

The Seebeck effect is a thermoelectric effect in which the voltage difference $V=(\muR-\muL)/e$ is brought about by the temperature gradient $\Delta T = \TR-\TL$ \cite{callen}. The Seebeck coefficient is defined by \cite{callen}
\begin{equation}
S \equiv -\left(\frac{V}{\Delta T}\right)_{\expect{J_N}=0}.
\end{equation}
Let us define the higher-order Seebeck coefficients as follows. Under the condition in which the average of the particle flow $\expect{J_N}$ is zero, the $k$th-order Seebeck coefficient is defined by \cite{sanchez2013,cipiloglu2004}
\begin{align}
-V &= S_1\Delta T + S_2(\Delta T)^2 + \cdots \notag \\
 &= \sum_{k=1}^{\infty}S_k(\Delta T)^k. \label{eq:nlSee}
\end{align}
Let us express the second-order Seebeck coefficient with the transport coefficients $G^{k_1,k_2}_{\ell_1,\ell_2}(B)$. 

In order to do this, we first express $\ANL$ in terms of $\AQL$ under the condition $\expect{J_N} = 0$, which is
 \begin{equation}
\expect{J_N} = G^{10}_{10}\ANL + G^{10}_{01}\AQL + \frac{G^{10}_{20}}{2!}\ANL^2+G^{10}_{11}\ANL\AQL + \frac{G^{10}_{02}}{2!}\AQL^2 =0.
\end{equation}
Solving this, we obtain two types of the affinity $\ANL$ as
\begin{equation}
\ANL_{\pm} = \frac{-G^{10}_{10}-G^{10}_{11}\AQL\pm\sqrt{(G^{10}_{10}+G^{10}_{11}\AQL^2)^2-G^{10}_{20}(2G^{10}_{01}\AQL+G^{10}_{02}\AQL^2)}}{G^{10}_{20}}, \label{eq:ANLpm}
\end{equation}
depending on the sign of the square root.
Note that we expanded $J_N$ only up to the second order of the affinities because the higher-order terms do not affect the result when we calculate the second-order Seebeck coefficient. 
Before expanding \eqref{eq:ANLpm}, we have to choose which solution we use. We choose $ A_{N+}^{\mathrm{L}}$ if $G^{10}_{01}>0$ and $ A_{N-}^{\mathrm{L}}$ if $G^{10}_{01}<0$ for the following two reasons. One reason is that the voltage would not be zero with $\Delta T =0$ unless we choose them in this way. 
A finite voltage with no particle flow and no temperature difference is not normal physically. The other reason is that the first-order Seebeck coefficient would not coincide with the well-known linear Seebeck coefficient \cite{callen}.

We then expand the properly chosen solution with respect to $\AQL$ to the second order:
\begin{equation}
\ANL = -\frac{G^{10}_{01}}{G^{10}_{10}}\AQL - \frac{G^{10}_{02}{G^{10}_{10}}^2-2G^{10}_{01}G^{10}_{10}G^{10}_{11}+{G^{10}_{01}}^2G^{10}_{20}}{2{G^{10}_{10}}^3} \AQL^2+O(\AQL^3), \label{eq:JN0ANL} 
\end{equation}
which turns out to be independent of the choice.
Inserting the affinities
\begin{align}
\AQL &= -\betaL+\betaR = \frac{\TL-\TR}{\TL\TR} = \frac{\Delta T}{\TL\TR}, \\
\ANL &= \betaR(\muL-\muR) = \frac{\muL-\muR}{\TR} = \frac{eV}{\TR},
\end{align}
for $\JQL$ and $J_N$ into Eq.~\eqref{eq:JN0ANL}, we have
\begin{equation}
\frac{eV}{\TR} = -\frac{G^{10}_{01}}{G^{10}_{10}} \frac{\Delta T}{\TL\TR} - \frac{G^{10}_{02}{G^{10}_{10}}^2-2G^{10}_{01}G^{10}_{10}G^{10}_{11}+{G^{10}_{01}}^2G^{10}_{20}}{2{G^{10}_{10}}^3} \left( \frac{\Delta T}{\TL\TR}\right)^2+O((\Delta T)^3), 
\end{equation}
which is followed by the expansion of the form \eqref{eq:nlSee}:
\begin{equation}
-V = \frac{G^{10}_{01}}{e\TL G^{10}_{10}}\Delta T + \frac{G^{10}_{02}{G^{10}_{10}}^2-2G^{10}_{01}G^{10}_{10}G^{10}_{11}+{G^{10}_{01}}^2G^{10}_{20}}{2e\TL^2\TR{G^{10}_{10}}^3}(\Delta T)^2.
\end{equation}
From this we obtain the first- and second-order Seebeck coefficients as follows:
\begin{align}
S_1 &=  \frac{1}{e\TL}\frac{G^{10}_{01}}{G^{10}_{10}}, \\
S_2 &=  \frac{1}{e\TL^2\TR}\frac{G^{10}_{02}{G^{10}_{10}}^2-2G^{10}_{01}G^{10}_{10}G^{10}_{11}+{G^{10}_{01}}^2G^{10}_{20}}{2{G^{10}_{10}}^3}. \label{eq:soSee}
\end{align}

Let us express the second-order Seebeck coefficient \eqref{eq:soSee} only with linear transport coefficients using the relation of the form \eqref{eq:mathLrela}. For simplicity, let us assume that there is no magnetic filed $B=0$.  We find from \eqref{eq:mathLrela} that the relations as \eqref{eq:L0110}--\eqref{eq:L0210-} is also valid for $G^{k_1,k_2}_{\ell 1,\ell 2}$:
\begin{align}
G^{10}_{20} &= G^{20}_{10}, \\
G^{11}_{10} &=2G^{10}_{11}-G^{20}_{01} \quad \Leftrightarrow \quad G^{10}_{11}=\frac{G^{11}_{10}+G^{20}_{01}}{2}, \\
G^{10}_{02} &= G^{11}_{01}, 
\end{align}
under no magnetic field. 
Using these relations, we can rewrite the second-order Seebeck coefficient \eqref{eq:soSee} as follows:
\begin{equation}
S_2 = \frac{1}{e\TL^2\TR}\frac{G^{11}_{01}{G^{10}_{10}}^2-G^{10}_{01}G^{10}_{10}(G^{11}_{10}+G^{20}_{01})+{G^{10}_{01}}^2G^{20}_{10}}{2{G^{10}_{10}}^3}. \label{eq:S_2}
\end{equation}
We thus expressed $S_2$ only with the linear-transport coefficients. 

In evaluating $S_2$ using the expression \eqref{eq:soSee}, it may be difficult to measure $G^{10}_{02}$ and $G^{10}_{20}$, which are nonlinear coefficients. Using the final expression \eqref{eq:S_2}, however, we can evaluate the nonlinear Seebeck coefficient $S_2$ by measuring the transport coefficients only in the linear-response regime. Indeed, it may seem difficult to observe the heat flow. Recently, however, the energy flow has been observed at low temperatures in the experiment \cite{jezouin2013}. We expect that the heat flow and its noise will become able to be observed experimentally in the future.

\chapter{Summary and Future Works}
In this thesis, we explained the details of the heat flow in mesoscopic one-dimensional transport systems and derived the generalized Onsager relations of flows which are superpositions of the particle and energy flows. 
In particular, we derived the relations among the pairs of $J_N$ and $\JQL$, $J_N$ and $\JQR$, and $\JQL$ and $\JQR$.

We here remark the use of the energy flow and the heat flow. In the research area of thermoelectricity, there seems to be a confusion in using the two flows. The concept of the heat flow becomes necessary when electrons do work, particularly when we consider a heat engine. We thus need to use the heat flow when we consider the work of electrons. 

The relations that we obtained here will be useful in the future. One possibility is to use them in calculating the nonlinear coefficients like the nonlinear Seebeck coefficient which we obtained in Chapter 4. 
In order to describe thermoelectric devices which are not in the linear-response regime, the conventional thermoelectric coefficients and the figure of merit, which are defined in the linear-response regime, should be insufficient. 
We thus need new coefficients which describe nonlinear thermoelectric effects. 
The relations among higher-order cumulants will be useful in evaluating the coefficients.  

Another possibility is the efficiency fluctuation. The research of the efficiency fluctuation appeared recently  \cite{verley2014a,verley2014b,polettini2014}. 
The authors in Ref.~\cite{verley2014a} proposed that when we consider the efficiency, we should replace the conventional second law
\begin{equation}
\expect{\Delta S_{\text{tot}}} \ge 0 \label{eq:secondlaw}
\end{equation}
with the fluctuation theorem
\begin{equation}
\expect{e^{- \Delta S_{\text{tot}}}} = 1, \label{eq:FT}
\end{equation}
where $\Delta S_{\text{tot}}$ is the total entropy production of the system. 
Combining Eq.~\eqref{eq:FT} with Jensen's inequality
\begin{equation}
\expect{e^{- \Delta S_{\text{tot}}}}\ge e^{-\expect{\Delta S_{\text{tot}}}},
\end{equation}
we can show that the fluctuation theorem \eqref{eq:FT} includes the second law \eqref{eq:secondlaw}; in other words, the fluctuation theorem is a higher entity than the second law.
It then necessitates us to consider the higher-order cumulants of the entropy production because the quantity $\expect{e^{- \Delta S_{\text{tot}}}}$ contains $\expect{ \Delta S_{\text{tot}}^2}$, $\expect{ \Delta S_{\text{tot}}^3}$, $\cdots$. The entropy production is expressed as in Eq.~\eqref{eq:SdotJ1J2} by the  generalized flows $J_1$ and $J_2$ defined in Eq.~\eqref{eq:J1J2}. We thus have to consider the higher-order cumulants of  the  generalized flows.
Using the generalized Onsager relations which we derived in Chapter 4, we may understand the behavior of the efficiency in the nonlinear-response regime.

The research of nonlinear thermoelectric devices, particularly as a heat engine, is in the dawn. We expect that the generalized Onsager relations that we derived in the present thesis will help the development of this research area and manufacturing high-efficiency thermoelectric devices using nonlinearity.

\appendix
\chapter{Detailed Calculations for Chapter 3}
In this appendix, we show the details of the calculations in Chapter 3.
\section{Calculation for Eq.~(\ref{eq:chara})}
In this section we show the details of calculation to obtain the relation \eqref{eq:chara}. We start from Eq.~\eqref{eq:ichicQn}, which we reproduce here:
\begin{equation}
Z(\{\chi_{cr}\},\{\chi_{hr}\};B) = \sum_{\{Q_{Nr}\},\{Q_{Er}\}}P(\{Q_{Nr}\},\{Q_{Er}\})\prod_{r=1}^{m}e^{i(\chi_{cr}Q_{Nr}+\chi_{hr}Q_{Er})}. 
\end{equation}
Inserting Eq.~\eqref{eq:3.27} into the above, we have
\begin{align} 
 Z(\{\chi_{cr}\},\{\chi_{hr}\};B) &= \sum_{\{Q_{Nr}\},\{Q_{Er}\}}\sum_{\nu,\lambda}P_{\nu\to \lambda}(\{Q_{Nr}\},\{Q_{Er}\})\braket{\Psi_\nu^{\rm{init}}|\rho_0|\Psi_\nu^{\rm{init}}} \notag \\
 &\quad \times \prod_{r=1}^{m}e^{i(\chi_{cr}Q_{Nr}+\chi_{hr}Q_{Er})}.
\end{align}
Using Eq.~\eqref{eq:3.26} further, we proceed as 
\begin{align}
 Z(\{\chi_{cr}\},\{\chi_{hr}\};B) &= \sum_{\{Q_{Nr}\},\{Q_{Er}\}}\sum_{\nu,\lambda} \left|\braket{\Psi_\lambda^{\rm{fin}}|e^{-iHs}|\Psi_\nu^{\rm{init}}}\right|^2 \braket{\Psi_\nu^{\rm{init}}|\rho_0|\Psi_\nu^{\rm{init}}} \notag \\
 &\quad \times \prod_{r=1}^{m}
 \left\{\delta[Q_{Nr} -(Q_{Nr,\nu}^{\rm{init}}-Q_{Nr,\lambda}^{\rm{fin}})]\delta[Q_{Er} -(Q_{Er,\nu}^{\rm{init}}-Q_{Er,\lambda}^{\rm{fin}})] \right. \notag \\
 &\quad \times \left.e^{i(\chi_{cr}Q_{Nr}+\chi_{hr}Q_{Er})}\right\} \notag \\
 &= \sum_{\nu,\lambda} \left|\braket{\Psi_\lambda^{\rm{fin}}|e^{-iHs}|\Psi_\nu^{\rm{init}}}\right|^2 \braket{\Psi_\nu^{\rm{init}}|\rho_0|\Psi_\nu^{\rm{init}}}\notag \\
 &\quad \times \prod_{r=1}^{m}e^{i[\chi_{cr}(Q_{Nr,\nu}^{\rm{init}}-Q_{Nr,\lambda}^{\rm{fin}})+\chi_{hr}(Q_{Er,\nu}^{\rm{init}}-Q_{Er,\lambda}^{\rm{fin}})]} \notag \\
 &= \sum_{\nu,\lambda} \braket{\Psi_\nu^{\rm{init}}|e^{iHs}\prod_{r=1}^{m}\left[e^{-i(\chi_{cr}Q_{Nr,\lambda}^{\rm{fin}}+\chi_{ch}Q_{Er,\lambda}^{\rm{fin}})}\right]|\Psi_\lambda^{\rm{fin}}} \notag \\
 &\quad \times \braket{\Psi_\lambda^{\rm{fin}}|e^{-iHs}\prod_{r=1}^{m}\left[e^{i(\chi_{cr}Q_{Nr,\nu}^{\rm{init}}+\chi_{ch}Q_{Er,\nu}^{\rm{init}})}\right]|\Psi_\nu^{\rm{init}}} \braket{\Psi_\nu^{\rm{init}}|\rho_0|\Psi_\nu^{\rm{init}}} \notag \\
 &= \sum_{\nu,\lambda} \braket{\Psi_\nu^{\rm{init}}|e^{iHs}\prod_{r=1}^{m}\left[e^{-i(\chi_{cr}N_r+\chi_{ch}H_r)}\right]|\Psi_\lambda^{\rm{fin}}} \notag \\
 &\quad \times \braket{\Psi_\lambda^{\rm{fin}}|e^{-iHs}\prod_{r=1}^{m}\left[e^{i(\chi_{cr}N_r+\chi_{ch}H_r)}\right]|\Psi_\nu^{\rm{init}}}  \braket{\Psi_\nu^{\rm{init}}|\rho_0|\Psi_\nu^{\rm{init}}} \notag \\
 &=  \sum_{\nu,\lambda} \braket{\Psi_\nu^{\rm{init}}|e^{iHs}V^2|\Psi_\lambda^{\rm{fin}}} \braket{\Psi_\lambda^{\rm{fin}}|e^{-iHs}{V^{\dagger}}^2|\Psi_\nu^{\rm{init}}}  \braket{\Psi_\nu^{\rm{init}}|\rho_0|\Psi_\nu^{\rm{init}}} \notag \\
 &= \Tr{e^{iHs}V^2e^{-iHs}{V^{\dagger}}^2\rho_0} \notag \\
 &=  \Tr{e^{iHs}V^2e^{-iHs}V^{\dagger}\rho_0 V^{\dagger}} \notag \\
 &= \Tr{\rho_0\dagg{V}e^{iHs}V^2e^{-iHs}\dagg{V}},
\end{align}
where
\begin{equation}
V = \prod_{r=1}^{m}\exp\left[\frac{-i(\chi_{cr} N_r+\chi_{hr}H_r)}{2}\right].
\end{equation}
We here used the relations $N_r\ket{\Psi_\nu^{\rm{init}}}=Q_{Nr,\nu}^{\rm{init}}\ket{\Psi_\nu^{\rm{init}}}$, $N_r\ket{\Psi_\lambda^{\rm{fin}}}=Q_{Nr,\lambda}^{\rm{fin}}\ket{\Psi_\lambda^{\rm{fin}}}$, $H_r\ket{\Psi_\nu^{\rm{init}}}=Q_{Er,\nu}^{\rm{init}}\ket{\Psi_\nu^{\rm{init}}}$, $H_r\ket{\Psi_\lambda^{\rm{fin}}}=Q_{Nr,\lambda}^{\rm{fin}}\ket{\Psi_\lambda^{\rm{fin}}}$, the fact that $\rho_0$ and $V^{\dagger}$ are commutative, and the invariance of the trace under cyclic permutations. 
\section{Calculation for Eq.~(\ref{eq:Fsymm})}
In this section, we prove the symmetry of the characteristic function 
\begin{equation}
Z(\{\chi_{cr}\},\{\chi_{hr}\};B) =Z(\{-\chi_{cr}+iA_{cr}\},\{-\chi_{hr}+iA_{hr}\};-B). \label{eq:AppZsymm}
\end{equation}

We first introduce the time-reversal operator $\Theta$ \cite{sakurai1985}. The operator $\Theta$ satisfies the following properties:
\begin{align}
\Theta i \Theta &= -i ,\\
\braket{n|O|n'} &= \braket{\tilde{n'}|\Theta O^{\dagger} \Theta^{\dagger}|\tilde{n}}, \\
\ket{\tilde{n}} &= \Theta\ket{n}, 
\end{align} 
where $\ket{n}$, $\ket{n'}$, and $O$ are bases and an operator in the Hilbert space, respectively.  
Let us then transform the left-hand side of Eq.~\eqref{eq:AppZsymm}. We here write the magnetic-field dependence of $H$ and $V$ explicitly. Note that $H(B)$ and $V(B)$ are transformed by the time-reversal operator as follows:
\begin{align}
\Theta H(B) \Theta &= H(-B), \\
\Theta V(B) \Theta &= V(-B).
\end{align}
We can therefore rewrite the characteristic function as follows:
\begin{align}
Z(\{\chi_{cr}\},\{\chi_{hr}\};B) &= \Tr{\rho_0\dagg{V(B)}e^{iH(B)s}V(B)^2e^{-iH(B)s}\dagg{V(B)}} \notag\\ 
 &= \sum_{n}\bra{n}\rho_0\dagg{V(B)}e^{iH(B)s}V(B)^2e^{-iH(B)s}\dagg{V(B)}\ket{n} \label{eq:first} \\
 &= \sum_{\tilde{n}}\bra{\tilde{n}}\Theta(\rho_0\dagg{V(B)}e^{iH(B)s}V(B)^2e^{-iH(B)s}\dagg{V(B)})^{\dagger}\Theta^{\dagger}\ket{\tilde{n}} \notag \\
 &= \sum_{\tilde{n}}\bra{\tilde{n}}\Theta(V(B)e^{iH(B)s}{V(B)^{\dagger}}^2e^{-iH(B)s}V(B)\rho_0)\Theta^{\dagger}\ket{\tilde{n}} \notag \\
 &= \sum_{\tilde{n}}\bra{\tilde{n}}\dagg{V(-B)}e^{-iH(-B)s}V(-B)^2e^{iH(-B)s}\dagg{V(-B)}\rho_0\ket{\tilde{n}} \notag \\
 &= \text{Tr}[e^{iH(-B)s}\dagg{V(-B)}\rho_0\dagg{V(-B)}e^{-iH(-B)s}V(-B)^2], \label{eq:last}
\end{align} 
where we assumed that the states $\{\ket{\tilde{n}}\}$ span a complete set if $\{\ket{n}\}$ do.

Because $H_r$ and $N_r$ commute with each other for any $r$, $\rho_0'$ defined in \eqref{eq:rho_0'}, or
\begin{equation}
\rho_0'(B) = \frac{1}{z}\prod_{r=1}^{m}e^{A_{hr}H_r(B)+A_{cr}N_r},
\end{equation} 
where $z=\prod_{r=1}^{m}\text{Tr}e^{A_{hr}H_r(B)+A_{cr}N_r}$ with $A_{hr}=-\beta_{r}$ and $A_{cr}=-\beta_{r}\mu_{r}$, and the operator $V$ defined in \eqref{eq:Vdef}, or more explicitly
\begin{equation}
V(\{\chi_{cr}\},\{\chi_{hr}\};B) = \prod_{r=1}^{m}\exp\left[\frac{-i}{2}(\chi_{cr}N_r+\chi_{hr}H_r(B)) \right],
\end{equation}
also commute with each other.
We can therefore write down the two terms in \eqref{eq:last} as follows: 
\begin{align}
\dagg{V(\{\chi_{cr}\},\{\chi_{hr}\};-B)}\rho_0'\dagg{V(\{\chi_{cr}\},\{\chi_{hr}\};-B)} &= \frac{1}{z}\prod_{r=1}^{m}e^{(A_{cr}+i\chi_{cr})N_r+(A_{hr}+i\chi_{hr})H_r(-B)}, \\
V(\{\chi_{cr}\},\{\chi_{hr}\};-B)^2 &= \prod_{r=1}^{m}e^{-i\chi_{cr}N_r-i\chi_{hr}H_r(-B)}.
\end{align}
After simple algebras, we can identify them with
\begin{align}
&\frac{1}{z}V(\{-\chi_{cr}+iA_{cr}\},\{-\chi_{hr}+iA_{hr}\};-B)^2, \\
&z\dagg{V(\{-\chi_{cr}+iA_{cr}\},\{-\chi_{hr}+iA_{hr}\};-B)}\rho_0'\dagg{V(\{-\chi_{cr}+iA_{cr}\},\{-\chi_{hr}+iA_{hr}\};-B)},
\end{align}
respectively. Equation \eqref{eq:last} is then rewritten as
\begin{align}
&\text{Tr}[e^{iH(-B)s}\dagg{V(-B)}\rho_0\dagg{V(-B)}e^{-iH(-B)s}V(-B)^2] \notag \\
&= Z(\{-\chi_{cr}+iA_{cr}\},\{-\chi_{hr}+iA_{hr}\};-B),
\end{align}
if $\rho_{d}$ commutes with $H(-B)$, where in the first line we left out the arguments $\{-\chi_{cr}+iA_{cr}\}$ and $\{-\chi_{hr}+iA_{hr}\}$ for brevity.

As is assumed after Eq.~\eqref{eq:rho_0'}, we here used $\rho_d = 1_d/2^n$, which indeed commutes with $H$. We thereby have the symmetry 
\begin{equation}
Z(\{\chi_{cr}\},\{\chi_{hr}\};B) =Z(\{-\chi_{cr}+iA_{cr}\},\{-\chi_{hr}+iA_{hr}\};-B).
\end{equation}

\chapter{The Sommerferd Expansion at Low Temperatures}\label{App4}

\section{The Sommerferd expansion at low temperatures}
In this appendix, we expand the heat flows $\JQL$ and $\JQR$ using the Sommerfeld expansion at low temperatures \cite{ashcroft1976}.
\subsection{The Sommerfeld expansion}
We can expand the integral
\begin{equation}
\int^{\infty}_{\EG}d\epsilon g(\epsilon)f(\epsilon) \label{eq:Sommer}
\end{equation}
in the form
\begin{equation}
\int^{\infty}_{\EG}d\epsilon g(\epsilon)f(\epsilon) = \int_{\EG}^{\mu}d\epsilon g(\epsilon) +\frac{\pi^2T^2}{6}g^{\prime}(\mu) + O(T^4),\label{eq:Sexpand}
\end{equation}
where $f(\epsilon)=\{1+\exp[\beta(\epsilon-\mu)]\}^{-1}$ is the Fermi distribution function and $g(\epsilon)$ is an arbitrary function which is continuous and infinitely differentiable with respect to $\epsilon$ at $\epsilon = \mu$, while diverges no more rapidly than some power of $\epsilon$ as $\epsilon \to +\infty$. Here, $g^{\prime}(\mu)$ denotes the first derivative of $g(\epsilon)$ with respect to $\epsilon$ at $\epsilon=\mu$.
The expansion (\ref{eq:Sexpand}) is called the Sommerfeld expansion \cite{ashcroft1976}.
\subsection{The derivation of the Sommerfeld expansion}
Let us derive the Sommerfeld expansion \eqref{eq:Sexpand}.
We first obtain the most rough approximation of the integral by setting $T=0$:
\begin{equation}
 \int_{\EG}^{\infty} d\epsilon \ g(\epsilon)f(\epsilon) \approx \int_{\EG}^{\mu} d\epsilon g(\epsilon). \label{eq:T=0}
\end{equation}
In order to approximate the integral  (\ref{eq:Sommer}) when the temperature is small but non-zero, we subtract the left-hand side from right-hand side of Eq.~(\ref{eq:T=0}):
\begin{align}
&\int_{\EG}^{\infty} d\epsilon g(\epsilon)f(\epsilon) - \int_{\EG}^{\mu} d\epsilon g(\epsilon) \\
&= \int_{\mu}^{\infty}d\epsilon g(\epsilon)f(\epsilon) - \int_{\EG}^{\mu}d\epsilon g(\epsilon)(1-f(\epsilon)) \\
&= \int_{\mu}^{\infty}d\epsilon \frac{g(\epsilon)}{1+e^{\beta(\epsilon-\mu)}} - \int_{\EG}^{\mu}d\epsilon \frac{g(\epsilon)}{1+e^{-\beta(\epsilon-\mu)}} \label{eq:3rdI} \\
&= T\int_{0}^{\infty}dx\frac{g(\mu+Tx)}{e^x+1} -T\int_{0}^{\beta(\mu-\EG)}dx\frac{g(\mu-Tx)}{e^x+1}. \label{eq:4thI}
\end{align}
We here used the transformation $x = \beta(\epsilon-\mu)$ in the first term and $x = -\beta(\epsilon-\mu)$ in the second term of Eq.~(\ref{eq:3rdI}).
Assuming $T\ll T_\mathrm{F}$, where $T_\mathrm{F}$ is the Fermi temperature defined as $T_\mathrm{F} \equiv \epsilon_\mathrm{F}$ where $\epsilon_\mathrm{F}$ is the Fermi energy, we have
\begin{equation}
\beta(\mu-\EG) = \frac{\mu-\EG}{T} \gg 1 ,
\end{equation}
and
\begin{equation}
\frac{1}{e^x+1} \ll1 \ \  \text{for}  \ \ x\gg1, 
\end{equation}
which show that the integrand of the second term in Eq.~(\ref{eq:4thI}) is exponentially small for large values of $x$. We can thus extend the upper limit of the integral of the second term in Eq.~(\ref{eq:4thI}) from $\beta(\epsilon-\mu)$ to infinity:
\begin{equation}
\int_{0}^{\beta(\mu-\EG)}dx\frac{g(\mu-Tx)}{e^x+1} \ \ \rightarrow \ \ \int_{0}^{\infty}dx\frac{g(\mu-Tx)}{e^x+1}.
\end{equation}
This lets us expand the integral $I$ in the form
\begin{align}
I &= T\int_{0}^{\infty}dx\frac{g(\mu+Tx)-g(\mu-Tx)}{e^x+1} \\
  &=  2g'(\mu)T^2\int_{0}^{\infty}dx\left(\frac{x}{e^x+1}\right) +O(T^4) \\
  &= \frac{\pi^2}{6}g'(\mu)T^2 + O(T^4),
\end{align}
where we used the Taylor expansion of $g$,
\begin{equation}
g(\mu\pm Tx) = g(\mu) \pm g'(\mu)Tx + \frac{g''(\mu)}{2}(Tx)^2 + \cdots,
\end{equation}
and the integral formula,
\begin{equation}
\int_{0}^{\infty}dx\frac{x}{e^x+1} = \frac{\pi^2}{12}.
\end{equation}
Note that 
\begin{equation}
g'(\mu) = \left. \frac{\mathrm{d}g(\epsilon)}{\mathrm{d}\epsilon}\right|_{\epsilon=\mu}.
\end{equation} 
\subsection{Expansion of the conventional heat flow}
We can expand the conventional heat flow using Eq.(\ref{eq:Sexpand}) as follows:
\begin{align}
\JQL &= \frac{1}{h} \int^{\infty}_{\EG}d\epsilon(\epsilon - \muL) (\fL(\epsilon)-\fR(\epsilon)) \\
       &= \frac{1}{h} \int^{\infty}_{\EG}d\epsilon(\epsilon - \muL)\fL(\epsilon) - \frac{1}{h} \int^{\infty}_{\EG}d\epsilon(\epsilon - \muL)\fR(\epsilon) \\
       &=  \frac{1}{h}\left[ \int_{\EG}^{\muL}d\epsilon (\epsilon - \muL)  +\frac{\pi^2\TL^2}{6} + O( \TL^4) \right] \nonumber \\
       & \ \ \  -\frac{1}{h}\left[ \int_{\EG}^{\muR}d\epsilon (\epsilon - \muL)  +\frac{\pi^2\TR^2}{6} + O(\TR^4) \right] \\
       &= \frac{\pi^2}{6h}(\TL^2-\TR^2) -\frac{1}{2h}(\muR-\muL)^2+ O(T^4),
\end{align}
\begin{align}
\JQR &= \frac{1}{h} \int^{\infty}_{\EG}d\epsilon(\epsilon - \muR) (\fL(\epsilon)-\fR(\epsilon)) \\
       &= \frac{1}{h} \int^{\infty}_{\EG}d\epsilon(\epsilon - \muR)\fL(\epsilon) - \frac{1}{h} \int^{\infty}_{\EG}d\epsilon(\epsilon - \muR)\fR(\epsilon) \\
       &=  \frac{1}{h}\left[ \int_{\EG}^{\muL}d\epsilon (\epsilon - \muR)  +\frac{\pi^2\TL^2}{6} + O(\TL^4) \right] \nonumber \\
       & \ \ \  -\frac{1}{h}\left[ \int_{\EG}^{\muR}d\epsilon (\epsilon - \muR)  +\frac{\pi^2\TR^2}{6} + O(\TR^4) \right] \\
       &= \frac{\pi^2}{6h}(\TL^2-\TR^2)  +\frac{1}{2h}(\muR-\muL)^2+ O(T^4) . \label{eq:jqrSexpand}
\end{align}

\chapter{The Upper Bound of The Efficiency of The Mesoscopic Heat Engine}\label{App3}
In this appendix, we show that the upper bound of the efficiency defined by (\ref{eq:effc}) is the Carnot efficiency and give an example of the transmission coefficient which achieves the upper bound. 
\section{The upper bound of the efficiency of the mesoscopic heat engine}
\subsection{General upper bound of the efficiency}
Let   $\SLdot$ and $\SRdot$ denote the entropy productions in the left and right reservoirs, respectively.
Using the equalities $\mathrm{d}Q = T\mathrm{d}S = \mathrm{d}U - \mu\mathrm{d}N$, we can relate these entropy productions to the conventional heat flows as 
\begin{align}
\JQL \equiv \dot{\UL}-\muL\dot{\NL} = -\TL\SLdot , \\
\JQR \equiv \dot{\UR}-\muR\dot{\NR} = \TR\SRdot.
\end{align}
These relations let us transform the efficiency \eqref{eq:effc} to the form
\begin{align}
\eta &= \frac{IV}{\JQL} \\ \label{eq:effc2}
       &= \frac{\JQL-\JQR}{\JQL} \\
       &= \frac{-\TL\SLdot -\TR\SRdot}{-\TL\SLdot} \\
       &= \frac{\TL\SLdot +\TR\SRdot}{\TL\SLdot}.
\end{align}
We here impose the condition that the net entropy production $\Sdot = \SLdot + \SRdot$ is positive, that is, $\SRdot \geq -\SLdot$, which gives 
\begin{align}
\eta &= \frac{\TL\SLdot +\TR\SRdot}{\TL\SLdot} \\
       &\leq \frac{\TL\SLdot -\TR\SLdot}{\TL\SLdot} \\
       &= 1-\frac{\TR}{\TL} = \eta_c,
\end{align}
where $\eta_c$ is the Carnot efficiency. We can achieve the equality if and only if $\SLdot =\SRdot$, that is, $\Sdot = 0$.
\subsection{Example of the transmission coefficient which gives the Carnot efficiency}
The Landauer-B\"uttiker formalism with the transmission coefficient gives
\begin{equation}
I = \frac{e}{h}\int_{\EG}^{\infty} d\epsilon \tau(\epsilon)(\fL(\epsilon)-\fR(\epsilon)),
\end{equation}
\begin{equation}
\JQL = \frac{1}{h}\int_{\EG}^{\infty} d\epsilon \tau(\epsilon)(\epsilon-\muL)(\fL(\epsilon)-\fR(\epsilon)).
\end{equation}
We thus express the efficiency (\ref{eq:effc2}) in the form
\begin{align}
\eta &= \frac{IV}{\JQL} \\
       &= \frac{(\muR-\muL)\int_{\EG}^{\infty} d\epsilon \tau(\epsilon)(\fL(\epsilon)-\fR(\epsilon))}{\int_{\EG}^{\infty} d\epsilon \tau(\epsilon)(\epsilon-\muL)(\fL(\epsilon)-\fR(\epsilon))}. \label{eq:eta}
\end{align}

We now know that we can achieve the Carnot efficiency when the total entropy production of the system is zero. We thus express the total entropy production with the Landauer-B\"uttiker formalism:
\begin{align}
\Sdot &= \SLdot + \SRdot \\
         &= \frac{\QL}{\TL} + \frac{\QR}{\TR} \\
         &= -\frac{\JQL}{\TL} + \frac{\JQR}{\TR} \\
         &= \int_{\EG}^{\infty} d\epsilon \tau(\epsilon)\left( -\frac{\epsilon-\muL}{\TL} + \frac{\epsilon-\muR}{\TR}\right)(\fL(\epsilon) - \fR(\epsilon)) \\
         &= \int_{\EG}^{\infty} d\epsilon \tau(\epsilon) \{[\log\fL(\epsilon)-\log(1-\fL(\epsilon))]-[\log\fR(\epsilon)-\log(1-\fR(\epsilon))]\} \nonumber \\
         &\times (\fL(\epsilon) - \fR(\epsilon)) \\
         &= \int_{\EG}^{\infty} d\epsilon \tau(\epsilon)(\fL(\epsilon) - \fR(\epsilon))\log\left[\frac{\fL(\epsilon)(1-\fR(\epsilon))}{\fR(\epsilon)(1-\fL(\epsilon))}\right]. \label{eq:sdotintegrand}
\end{align} 
Let us assume $\tau(\epsilon) \ge 0$, which is physically reasonable. We can show that the integrand in Eq.~(\ref{eq:sdotintegrand}) is always non-negative as follows. When $\fL(\epsilon) \ge \fR(\epsilon)$, 
\begin{align}
\fL(\epsilon)(1-\fR(\epsilon)) &= \fL(\epsilon) - \fL(\epsilon)\fR(\epsilon) \\
	&\ge \fR(\epsilon) - \fL(\epsilon)\fR(\epsilon) \\
	&= \fR(\epsilon)(1-\fL(\epsilon)),
\end{align}
which leads to 
\begin{equation}
\log\left[\frac{\fL(\epsilon)(1-\fR(\epsilon))}{\fR(\epsilon)(1-\fL(\epsilon))}\right] \ge 0.
\end{equation}
When $\fL(\epsilon) \le \fR(\epsilon)$, on the other hand, we can show 
\begin{equation}
\log\left[\frac{\fL(\epsilon)(1-\fR(\epsilon))}{\fR(\epsilon)(1-\fL(\epsilon))}\right] \ge 0
\end{equation} 
similarly. 
Using these inequalities and $\tau(\epsilon) \ge 0$, we can show that the integrand is always non-negative:
\begin{equation}
\tau(\epsilon)(\fL(\epsilon) - \fR(\epsilon))\log\left[\frac{\fL(\epsilon)(1-\fR(\epsilon))}{\fR(\epsilon)(1-\fL(\epsilon))}\right] \ge 0.
\end{equation}

Let us then consider the condition for $\Sdot = 0$. We easily find the following condition; for each value of $\epsilon$, $\tau(\epsilon)=0$ or $\fL(\epsilon) - \fR(\epsilon) = 0$. If $\tau(\epsilon)=0$ for any $\epsilon$ or if $\fL(\epsilon) - \fR(\epsilon) = 0$ for any $\epsilon$, however, the transport would not happen, that is, the condition is trivial. We thus have to find a nontrivial condition in which $\tau(\epsilon) \neq 0$ and  $\fL(\epsilon) - \fR(\epsilon) = 0$ at an energy; if we demanded $\fL(\epsilon) = \fR(\epsilon)$ at two energies, they would be equal at any energy.  

In order to do this, let us set the transmission function \cite{humphrey2002}
\begin{equation}
\tau(\epsilon) = \delta(\epsilon - \epsilon_c), \label{eq:tau}
\end{equation}
where $\epsilon_c$ is a constant. This transmission function is not a sufficient condition but a necessary condition for us to obtain $\Sdot = 0$. Let us also set the condition
 \begin{equation}
\fL(\epsilon_c) = \fR(\epsilon_c),
\end{equation}
which gives the value of $\epsilon_c$ as follows:
\begin{align}
&\fL(\epsilon_c) = \fR(\epsilon_c) \\
&\Leftrightarrow \quad \frac{1}{1+e^{\betaL(\epsilon_c-\muL)}} = \frac{1}{1+e^{\betaR(\epsilon_c-\muR)}} \\
&\Leftrightarrow \quad \betaL(\epsilon_c-\muL) = \betaR(\epsilon_c-\muR) \\
&\Leftrightarrow \quad \epsilon_c = \frac{\TL\muR-\TR\muL}{\TL-\TR}. \label{eq:epsilonc}
\end{align} 
Substituting Eq.~(\ref{eq:tau}) into Eq.~(\ref{eq:eta}), we obtain the following expression of the efficiency:
\begin{align}
\eta &= \frac{(\muR-\muL)\int_{\EG}^{\infty} d\epsilon \tau(\epsilon)(\fL(\epsilon)-\fR(\epsilon))}{\int_{\EG}^{\infty} d\epsilon \tau(\epsilon)(\epsilon-\muL)(\fL(\epsilon)-\fR(\epsilon))} \\
       &= \frac{(\muL-\muR)(\fL(\epsilon_c)-\fR(\epsilon_c))}{(\epsilon_c-\muL)(\fL(\epsilon_c)-\fR(\epsilon_c))} \\
       &=  \frac{\muL-\muR}{\epsilon_c-\muL}. \label{eq:etac}
\end{align}

Substituting Eq.~(\ref{eq:epsilonc}) into Eq.~(\ref{eq:etac}), we indeed achieve the Carnot efficiency:
\begin{align}
\eta &= \frac{\muL-\muR}{\epsilon_c-\muL} \\
       &= \frac{\muL-\muR}{ \frac{\TL\muR-\TR\muL}{\TL-\TR}-\muL} \\
       &= \frac{\TL\muR-\TL\muL-\TR\muR+\TR\muL}{\TL\muR-\TL\muL} \\
       &= \frac{\TL(\muR-\muL)-\TR(\muR-\muL)}{\TL(\muR-\muL)} \\
       &= 1-\frac{\TR}{\TL} = \eta_c.
\end{align}
Note that when $\eta=\eta_c$, the electric current and the heat current vanish:
\begin{align}
I &= \frac{e}{h}\int_{\EG}^{\infty} d\epsilon \tau(\epsilon)(\fL(\epsilon)-\fR(\epsilon)) \\
  &=  \frac{e}{h}(\fL(\epsilon_c)-\fR(\epsilon_c))  \ \ (\because \tau(\epsilon) = \delta(\epsilon - \epsilon_c))\\
  &= 0 \ \ ( \because \fL(\epsilon_c) = \fR(\epsilon_c)),
\end{align}
\begin{align}
\JQL &= \frac{1}{h}\int_{\EG}^{\infty} d\epsilon \tau(\epsilon)(\epsilon-\muL)(\fL(\epsilon)-\fR(\epsilon)) \\
  &=  \frac{1}{h}(\epsilon_c-\muL)(\fL(\epsilon_c)-\fR(\epsilon_c)) \ \ (\because \tau(\epsilon) = \delta(\epsilon - \epsilon_c))\\
  &= 0 \ \ ( \because \fL(\epsilon_c) = \fR(\epsilon_c)),
\end{align}
which gives the vanishing power $IV$. This is the same as the standard heat engine; the Carnot cycle produces zero power. 




\begin{thebibliography}{99}

\bibitem{goupil2011}
C. Goupil, W. Seifert, K. Zabrocki, E. M\"uller, and G.~J.~ Snyder, Entropy \textbf{13}, 1481 (2011).

\bibitem{callen}
H.~B.~Callen, \textit{Thermodynamics and an Introduction to Thermostatistics} (2nd Ed.), Wiley, New York (1985).

\bibitem{onsager1931a}
L. Onsager, Phys. Rev. \textbf{37}, 405 (1931).

\bibitem{onsager1931b}
L. Onsager, Phys. Rev. \textbf{38}, 2265 (1931).

\bibitem{casimir1945}
H.~B.~G.~Casimir, Rev. Mod. Phys. \textbf{17}, 343 (1945).

\bibitem{vandenboreck2005}
C. Van den Broeck, Phys. Rev. Lett. \textbf{95}, 190602 (2005). 

\bibitem{saito2010}
K.~Saito, G.~Benenti, and G.~Casati, Chem. Phys. \textbf{375}, 508 (2010).

\bibitem{benenti2011}
G.~Benenti, K.~Saito, and G.~Casati, Phys. Rev. Lett. \textbf{106}, 230602 (2011).

\bibitem{saito2011}
K. Saito, G. Benenti, G. Casati, and T. Prosen, Phys. Rev. B \textbf{84}, 201306(R) (2011).

\bibitem{brandner2013}
K.~Brandner, K.~Saito, and U.~Seifert, Phys. Rev. Lett. \textbf{110},  070603 (2013).

\bibitem{brandner2013multi}
K.~Brandner and U.~Seifert, New J. Phys. \textbf{15}, 105003 (2013).

\bibitem{Benenti}
G.~Benenti, G.~Casati, T.~Prozen, and K.~Saito, arXiv:1311.4430 (2013).

\bibitem{landauer1957}
R. Landauer, IBM J. Res. \& Dev. \textbf{1}, 223 (1957).

\bibitem{datta1995}
S. Datta, \textit{Electronic Transport in Mesoscopic Systems}, Cambridge University Press, Cambridge, UK (1995).

\bibitem{diventra2008}
M.~Di Ventra, \textit{Electrical Transport in Nanoscale Systems}, Cambridge University Press, Cambridge, UK, (2008).

\bibitem{kato2014}
T. Kato, Bussei Kenkyu Denshiban \textbf{3}, 1 (2014) (in Japanese).

\bibitem{wees1988}
B. J. van Wees, H. van Houten, C. W. J. Beenakker, J. G. Williamson, L. P. Kouwenhoven, D. van der Marel, and C. T. Foxon, Phys. Rev. Lett. \textbf{60}, 848 (1988).
 
 \bibitem{snyder2008}
G.~J.~Snyder, E.~S.~Toberer, Nat. Mater. \textbf{7}, 105 (2008).

\bibitem{matthews2014}
J. Matthews, F. Battista, D. S\'anchez, P. Samuelsson, and H. Linke, Phys. Rev. B \textbf{90}, 165428 (2014).

\bibitem{izumida2012}
Y. Izumida and K. Okuda, Europhys. Lett. \textbf{97}, 10004 (2012).

\bibitem{izumida2014}
Y. Izumida, K. Okuda, J.~M.~M.~Roco, and A.~C.~Hernandez, arXiv:1405.6777 (2014).

\bibitem{saito2008}
K. Saito and Y. Utsumi, Phys. Rev. B \textbf{78}, 115429 (2008).

\bibitem{nazarov2003}
Yu.~V.~Nazarov (ed.),\textit{Quantum Noise in Mesoscopic Physics}, NATO Science Series II: Mathematics, Physics and Chemistry, Vol. \textbf{97},  Kluwer Academic, Dordrecht (2003).

\bibitem{saito2009}
K. Saito, Bussei Kenkyu, \textbf{92}, 345 (2009) (in Japanese).

\bibitem{martin2005}
T. Martin, arXiv:cond-mat/0501208 (2005).

\bibitem{fetter2003}
A.~L.~Fetter and J.~D.~Walecka, \textit{Quantum Theory of ManyParticle Systems}, Dover Publications, Inc, New York (2003).

\bibitem{kurchan2001}
J. Kurchan, arXiv:cond-mat/0007360 (2001).

\bibitem{nakamura2010}
S. Nakamura, Y. Yamauchi, M. Hashisaka, K. Chida, K. Kobayashi, T. Ono, R. Leturcq, K. Ensslin, K. Saito, Y. Utsumi, and A.~C.~Gossard, Phys. Rev. Lett. \textbf{104}, 080602 (2010).

\bibitem{nakamura2011}
S. Nakamura, Y. Yamauchi, M. Hashisaka, K. Chida, K. Kobayashi, T. Ono, R. Leturcq, K. Ensslin, K. Saito, Y. Utsumi, and A.~C.~Gossard, Phys. Rev. B \textbf{83}, 155431 (2011).

\bibitem{iyoda2010}
E. Iyoda, Y. Utsumi, and T. Kato, J. Phys. Soc. Jpn. \textbf{79}, 045003 (2010).

\bibitem{ruokola2011}
T. Ruokola and T. Ojanen, Phys. Rev. B \textbf{83}, 241404(R) (2011).

\bibitem{lim2013}
J.~S.~Lim, R.~L\'opez, and D.~S\'anchez, Phys. Rev. B \textbf{88}, 201304(R) (2013).

\bibitem{sivan}
U.~Sivan and Y. Imry, Phys. Rev. B \textbf{33}, 551 (1986).

\bibitem{Butcher}
P.~N.~Butcher, J. Phys. Condens. Matter \textbf{2}, 4869 (1990).

\bibitem{jezouin2013}
S.~Jezouin, F.~D.~Parmentier, A.~Anthore, U.~Gennser, A.~Cavanna, Y.~Jin, and F.~Pierre, Science \textbf{342}, 601 (2013).

\bibitem{iyoda}
E.~Iyoda, \textit{Thermoelectric effects and fluctuation theorem in quantum dots}, Master Thesis, University of Tokyo (2008) (in Japanese); Bussei Kenkyu \textbf{92}, 188 (2009) (in Japanese).

\bibitem{sanchez2013}
D.~S\'anchez and R.~L\'opez, Phys. Rev. Lett. \textbf{110},  026804 (2013).

\bibitem{cipiloglu2004}
M.~A.~{\c{C}}ipilo{\u{g}}lu, S.~Turgut, and M.~Tomak, Phys. Status Solidi B \textbf{241}, 2575 (2004). 

\bibitem{verley2014a}
G. Verley, M. Esposito, T. Willaert, and C. Van den Broeck, Nat. Commun. \textbf{5}, 4721 (2014). 
 
\bibitem{verley2014b}
G. Verley, T. Willaert, C. Van den Broeck, and M. Esposito, Phys. Rev. E \textbf{90}, 052145 (2014).
 
\bibitem{polettini2014}
M. Polettini, G. Verley, and M. Esposito, arXiv:1409.4716 (2014).

\bibitem{sakurai1985}
J.~J.~Sakurai, \textit{Modern Quantum Mechanics}, Benjamin, Menlo Park, California (1985).

\bibitem{ashcroft1976}
N.~W.~Ashcroft and N.~D.~Mermin, \textit{Solid State Physics}, Holt, Rinehart and Winston, New York (1976).

\bibitem{humphrey2002}
T.~E.~Humphrey, R. Newbury, R.~P.~Taylor, and H. Linke, Phys. Rev. Lett. \textbf{89}, 116801 (2002).

\end{thebibliography}
\end{document}